\documentclass[preprint,superscriptaddress,nofootinbib]{revtex4}

\usepackage{graphicx}
\usepackage{dcolumn}
\usepackage{amsmath}
\usepackage[colorlinks=true, citecolor=blue, urlcolor = blue, linkcolor= red, bookmarks=true]{hyperref}

\begin{document}
\title{Rotating five-dimensional electrically charged Bardeen regular black holes}

\author{Muhammed Amir}
\email{amirctp12@gmail.com}
\affiliation{Astrophysics and Cosmology Research Unit, School of Mathematics, 
Statistics and Computer Science, University of KwaZulu-Natal, 
Private Bag X54001, Durban 4000, South Africa}
\author{Md Sabir Ali} 
\email{sabir.ali@iitrpr.ac.in} 
\affiliation{Indian Institute of Technology Ropar, Punjab-140001, India}
\author{Sunil D. Maharaj}
\email{maharaj@ukzn.ac.za}
\affiliation{Astrophysics and Cosmology Research Unit, School of Mathematics, 
Statistics and Computer Science, University of KwaZulu-Natal, 
Private Bag X54001, Durban 4000, South Africa}

\begin{abstract}
We derive a rotating counterpart of the five-dimensional electrically charged Bardeen 
regular black holes spacetime by employing the Giampieri algorithm on static one. 
The associated nonlinear electrodynamics source is computed in order to justify the 
rotating solution. We thoroughly discuss the energy conditions and the other properties 
of the rotating spacetime. The black hole thermodynamics of the rotating spacetime is 
also presented. In particular, the thermodynamic quantities such as the Hawking 
temperature and the heat capacity are calculated and plotted to see the thermal 
behavior. The Hawking temperature profile of the black hole implies that the regular 
black holes are thermally colder than its singular counterpart. On the other hand, we 
find that the heat capacity has two branches: the negative branch corresponds to the 
unstable phase and the positive branch corresponds to that of the stable phase for a 
suitable choice of the physical parameters characterizing the black holes. 
\end{abstract}
 
\maketitle

\section{Introduction}
The existence of the equivalence principle and the connection between gravity and 
thermodynamics are two of the most fascinating features of gravity. Between these two, 
the principle of equivalence finds its natural expression when gravity is described as 
a manifestation of curved spacetime \cite{Padmanabhan:2002ma}. Black holes are the 
automatic consequence of the Einstein's field equations. The first exact mathematical 
solution of black holes was discovered by Karl Schwarzschild widely known as the 
Schwarzschild solution \cite{Schwarzschild:1916}, which describes a static spherically 
symmetric spacetime. However, the charged version of the Schwarzschild solution is known 
as the Reissner-Nordstr{\"o}m solution \cite{Reissner:1916}. After the proposal 
of these spacetimes various solutions of black holes have been discovered. Black holes 
have not been only studied in four dimensions but also there are many higher dimensional 
solutions available in the literature. The observational detection of gravitational 
waves by advanced LIGO/Virgo is strong confirmation of the existence of black holes 
\cite{Abbott:2016blz,Abbott:2016nmj,TheLIGOScientific:2016pea}. Recent observations by 
the Event Horizon Telescope revealed the first image of suppermassive black hole, 
namely M87 \cite{Akiyama:2019fyp,Akiyama:2019eap}.

Finding exact solutions of the Einstein's field equations is a nontrivial task and a 
very important branch of general relativity. In general relativity, it is always 
challenging to discover an axisymmetric or rotating solution in four dimensions as well 
as in higher dimensions. Roy Kerr obtained the well-known axisymmetric solution 
\cite{Kerr:1963ud}; it was discovered by imposing the stationary and axial symmetric 
conditions on the Einstein's field equations \cite{Kerr:1963ud}. Physically, these 
rotating solutions have paramount importance because it is believed that most of the 
astrophysical black holes are rotating or Kerr-like. In 1965, Newman and Janis 
discovered an alternative approach to obtain the same Kerr metric by imposing a set 
of complex coordinate transformations on the Schwarzschild  metric 
\cite{Newman:1965tw}. Through these transformations angular momenta are added to 
the static seed metric; this is widely known as the Newman Janis Algorithm (NJA). 
Furthermore, an alternative derivation of the Kerr-Newman metric is derived from the 
Reissner-Nordstr{\"o}m seed metric by using the same algorithm \cite{Newman:1965my}. 
There exists a wide range of rotating solutions, discovered through this algorithm 
\cite{Demianski:1966,Demianski:1972uza,Herrera:1982,Erbin:2016lzq,Kim:1998iw,Yazadjiev:1999ce}. 
The NJA approach has also been used to describe the Kerr interiors from static 
metrices \cite{Drake:1997hh}. Apart from NJA, several other methods have been 
developed to generate the rotating solutions from the static ones
\cite{Clement:1997tx,Drake:1998gf,Glass:2004rr,Gibbons:2004uw,Kim:2012pb,Murenbeeld:1970aq}.
Giampieri \cite{Giampieri:1990} proposed another alternative algorithm to generate the 
rotating metric in a much more simple way than the NJA. This algorithm requires neither 
calculation of the null tetrad basis nor is the metric to be inverted. It is noticeable 
that the results discovered through both of these techniques are exactly same. 
Recently, this algorithm was used by the authors in 
\cite{Erbin:2014lwa,Mirzaiyan:2017adt} to obtain successfully the five-dimensional 
Myers-Perry solution \cite{Myers:1986un} from the static Schwarzschild-Tangherlini 
seed metric. Since the null tetrad calculation is very tedious while working in 
higher dimensions \cite{Myers:1986un,Dianyan:1988}, this algorithm may 
be very helpful in extending to higher dimensions. 

Our main motive of this paper is to apply the Giampieri algorithm on the five-dimensional 
electrically charged static Bardeen regular spacetime in order to obtain a rotating 
spacetime having two distinct spin parameters. Moreover, our main motivation in th work is to discuss 
the various properties of the rotating spacetime, e.g., horizons, static limit surfaces, 
and ergosphere. As we already know that the Bardeen spacetime represents a singularity 
free black hole spacetime. It is a general belief that black holes contain spacetime 
singularities and the cosmic censorship conjecture claims that these spacetime 
singularities are covered by the event horizon. Their presence breaks down the 
standard physical laws and hence it provides limitations to general relativity. 
Quantum gravity is the expected theory that might be capable to resolve the singularity 
problem. This theory discusses very tiny scale physics (at the Planck scale), but we are 
still far away from achieving this. It is required that we consider some alternative 
approaches that are somehow able to remove or to avoid these spacetime singularities. 
One of the possible approach to obtain such nonsingular nature of the spacetime is the prediction of the regular black hole solutions where a central singularity is replaced by a de-Sitter belt. These 
solutions contain an event horizon but not any spacetime singularity within them. The 
spacetime singularities are avoided either by introducing some exotic field in the 
form of nonlinear electrodynamics or by some modification to gravity. Bardeen applied 
the idea of Sakharov \cite{Sakharov:1966} to replace the spacetime singularity by a 
de Sitter core, and described the first black hole solution without any spacetime 
singularity \cite{Bardeen:1968}. The physical source associated with Bardeen black hole 
was described in \cite{AyonBeato:2000zs}. The rotating Bardeen solution is proposed in 
\cite{Bambi:2013ufa}, in which the authors applied NJA on the Bardeen solution 
\cite{Bardeen:1968}. After the discovery of the Bardeen solution several regular 
static and rotating black hole solutions has been published in literature, which 
successively overcome spacetime singularities 
\cite{AyonBeato:1998ub,AyonBeato:1999ec,Dymnikova:2004zc,Bronnikov:2000vy,Shankaranarayanan:2003qm,Hayward:2005gi,Culetu:2014lca,Balart:2014cga,Toshmatov:2014nya,Ghosh:2014hea,Neves:2014aba,Azreg-Ainou:2014nra,Azreg-Ainou:2014pra,Ghosh:2014pba} .
Recently, the $d$-dimensional Bardeen-de Sitter black holes are presented 
\cite{Ali:2018boy} and regular black holes in Einstein-Gauss-Bonnet gravity are also 
obtained \cite{Ghosh:2018bxg,Kumar:2018vsm}.

The paper is organized as follows. We discuss the five-dimensional electrically charged 
Bardeen regular black hole spacetime in Sec.~\ref{solution}. In Sec.~\ref{giamp}, 
we employ the Giampieri algorithm on the static metric and obtain a rotating 
five-dimensional electrically charged Bardeen regular spacetime. The energy conditions 
of the rotating spacetime are analysed in Sec.~\ref{enc} and the nature of the spacetime 
in Sec.~\ref{prop}. Black hole thermodynamics is the subject of Sec.~\ref{bhthd}. We 
conclude by discussing our main results in Sec.~\ref{conclusion}.

\section{Electrically charged Bardeen regular black holes} 
\label{solution}
The dynamics of the spacetime is governed by the action, therefore, the 
Einstein-Hilbert action coupled to the nonlinear electrodynamics in five 
dimensions can be written as follows
\begin{equation}\label{action}
   I = \frac{1}{16\pi}\int{d^5{x}\sqrt{-g}\left[R-4\mathcal{L}
    \left(\mathcal{F}\right)\right]},
\end{equation}
where $R$ is the Ricci scalar and $g$ is related to the determinant of spacetime 
metric. The Lagrangian $\mathcal{L}(\mathcal{F})$ is a nonlinear function of the 
electromagnetic field strength $\mathcal{F} = F_{\mu\nu} F^{\mu\nu}/4$ where 
$F_{\mu\nu}=\partial_{\mu}A_{\nu} -\partial_{\nu}A_{\mu}$, and $A_{\mu}$ is 
the gauge potential. The Einstein's field equations  can be derived by the variation 
of action (\ref{action}) with respect to $g_{\mu \nu}$, which yields
\begin{eqnarray}\label{einstein}
R_{\mu\nu}-\frac{1}{2}g_{\mu\nu}R &=& T_{\mu\nu}.
\end{eqnarray}
However, the energy-momentum tensor $T_{\mu\nu}$ is given by
\begin{eqnarray}\label{einstein2}
T_{\mu\nu}=2\left(\mathcal{L}_{\mathcal{F}}F_{\mu\alpha}F_{\nu}^{\alpha}
-g_{\mu\nu}\mathcal{L}\right),
\end{eqnarray}
where $\mathcal{L}_{\mathcal{F}}=\partial\mathcal{L}/\partial\mathcal{F}$. 
On the other hand, the Maxwell equations can be obtained by varying the action with 
respect to gauge potential $A_{\mu}$, which read simply
\begin{eqnarray}\label{maxwell}
\nabla_{\mu}\left(\mathcal{L}_\mathcal{F}F^{\mu\nu}\right) &=& 0.
\end{eqnarray}
The static spacetime metric of the five-dimensional electrically charged 
Bardeen regular black hole \cite{Ali:2018boy} is given by
\begin{eqnarray}\label{metric}
ds^2 &=& -f(r) dt^2 + f(r)^{-1} dr^2 +r^2 \left(d\theta^2+\sin^2\theta d\phi^2
+\cos^2\theta d\psi^2\right),
\end{eqnarray}
with 
\begin{equation}
 f(r) =  1 - \frac{\mu r^2}{(r^3 +q_e^3)^{4/3}},
\end{equation}
where the free parameters $q_e$ and $\mu$ are related to the electric charge and 
black hole mass, respectively. The metric function $f(r)$ for large and small $r$, 
respectively, reads
\begin{eqnarray}\label{asymp}
f(r)\sim 1-\frac{\mu}{r^{2}}, \quad f(r) \sim 1-\frac{\mu}{q_e^4} r^{2}.
\end{eqnarray}

In order to obtain the electromagnetic tensor, we solve the Maxwell 
equations (\ref{maxwell}). The four dimensional solution of electrically 
charged Bardeen black hole has been discussed by Rodrigues {\it et al.}
\cite{Rodrigues:2018bdc}. The non-vanishing components of the 
electromagnetic tensor are $F_{01}$ and $F_{10}$, which read simply
\begin{equation}
    F_{01} = \frac{q_e}{\mathcal{L}_{\mathcal{F}}(r) r^3}= -F_{10}.
\end{equation}
Therefore, the electromagnetic field strength takes the following form,
\begin{equation}\label{scaf}
    \mathcal{F} = - \frac{q_e^2}{2 \mathcal{L}_\mathcal{F}^2(r) r^6 }.
\end{equation}
We further solve the Einstein's field equations in order to obtain the 
expressions of $\mathcal{L}(r)$ and $\mathcal{L}_\mathcal{F}(r)$, as
\begin{eqnarray}\label{sour}
\mathcal{L}(r) &=& \frac{\mu q_e^3 \left(3q_e^3 -4r^3\right)}
{\left(r^3 +q_e^3\right)^{10/3}},
\nonumber \\
\mathcal{L}_{\mathcal{F}}(r) &=&  \frac{\left(r^3 +q_e^3\right)^{10/3}}
{7 \mu q_e r^9}.
\end{eqnarray}
These equations when substitute back into the Einstein field equations, 
they satisfy them. The corresponding gauge potential can be expressed as following
\begin{equation}\label{gauge}
 A^{\mu}= -\frac{\mu r^7}{q_e \left(r^3 + q_e^3\right)^{7/3}}
 \delta^{\mu}_t.
\end{equation}
A substitution of $\mathcal{L}_\mathcal{F}(r)$ from (\ref{sour}) into 
(\ref{scaf}) yields
\begin{equation}\label{nfs}
    \mathcal{F} = - \frac{49 \mu^2 q_e^4 r^{12}}
    {2 \left(r^3 +q_e^3\right)^{20/3}}.
\end{equation}
In order to verify that the spacetime (\ref{metric}) does not contain any spacetime 
singularity, we need to obtain the expressions of the curvature invariants, 
and check whether they diverge or give finite value in the limit $r\rightarrow 0$. 
We can compute these limits as follows
\begin{eqnarray}
\lim_{r \rightarrow 0} R = \frac{20 \mu}{q_e^4}, \quad 
\lim_{r \rightarrow 0} \mathcal{R} = \frac{80 \mu^2}{q_e^8}, \quad
\lim_{r \rightarrow 0} K = \frac{40 \mu^2}{q_e^8}.
\end{eqnarray}
It is clear that the black hole interior does not terminate in a singularity but 
crosses the Cauchy horizon and develops in a region that becomes more and more 
de-Sitter like and eventually ending with a regular origin at $r=0$. Now let us step 
forward to achieve our goal to describe the rotating counterpart of this solution.
 
\section{Rotating five-dimensional electrically charged Bardeen regular 
black holes}
\label{giamp}
In this section, we formulate the rotating counterpart of the 
five-dimensional electrically charged Bardeen regular spacetime by employing the 
Giampieri algorithm on the static metric. The Giampieri algorithm is much simpler 
algorithm to generate the rotating solution in comparison to the Newman-Janis algorithm 
(NJA). A very simple reason is of course this algorithm avoids the tedious null tetrads 
calculation. The important feature that we encounter in NJA is the fact that a given 
set of transformations in $(r,\phi)$-plane generates rotation in the later. The 
generation of two different angular momenta in two different planes would then require 
successive applications of the NJA on different hypersurfaces. In five dimensions, 
the two different planes that can be made rotating are the $(r,\phi)$ and $(r,\psi)$ 
planes. In order to do that we need to dissociate the radii of the two planes in order 
to apply the NJA on each plane and thereby generating two different angular momenta in 
respective planes. The important thing about this prescription is that 
we employ it to generate five-dimensional doubly rotating spacetime. 
We must dissociate the parts of the metric that correspond to the rotating and 
nonrotating two-planes. Therefore, one has to take care of the $r^2$-term from 
transforming under complex transformation in the part of the metric 
defining the plane which will stay static.

Let us move forward step by step to construct the rotating five-dimensional electrically 
charged Bardeen regular black hole spacetime. The first step of the Giampieri algorithm 
is similar to that of NJA where we need to write down the metric (\ref{metric}) in the 
Edington-Finklestein retarded null coordinates. In addition, we introduce a function 
$R(r)=Re(r)$ such that the metric in null coordinates is written 
\cite{Erbin:2014lwa} as follows
\begin{eqnarray}\label{metric1}
ds^2=-du(du+2dr)+(1-f(r))du^2+r^2\left(d\theta^2+\sin^2\theta d\phi^2\right)
+R^2\cos^2\theta d\psi^2.
\end{eqnarray}
The coordinates $u$ and $r$ must be complex and the metric (\ref{metric1}) transformed 
in $(r, \phi)$-plane by the following coordinate transformations
\begin{eqnarray}\label{complex1}
u &=& u^{\prime} +ia\cos\chi_1,\quad r=r^{\prime} -ia\cos\chi_1,
\end{eqnarray}
where $\chi_1$ is a new angle. The differential transformations of (\ref{complex1})  
are given by
\begin{eqnarray}\label{complex01}
du &=& du^{\prime} -a\sin^2\theta d\phi, \quad dr=dr^{\prime} 
+a\sin^2\theta d\phi,
\end{eqnarray}
where we use the ansatz $i d\chi_1 = \sin\chi_1 d\phi$ and $\chi_1=\theta$. On 
applying the complex transformations (\ref{complex1}) and (\ref{complex01}) on metric 
(\ref{metric1}) as well as by replacing the metric function $f(r)$ with 
$\tilde{f}^{\lbrace1\rbrace}=\tilde{f}^{\lbrace1\rbrace}(r,\theta)$, and after omitting 
the primes, we obtain 
\begin{eqnarray}\label{metric2}
ds^2 &=& -du^2-2dudr+\left(1-\tilde{f}^{\lbrace1\rbrace}\right)
\left(du-a\sin^2\theta d\phi\right)^2+2a\sin^2\theta dr d\phi \nonumber\\ 
&& +(r^2+a^2\cos^2\theta)d\theta^2 +(r^2+a^2)\sin^2\theta d\phi^2
+r^2\cos^2\theta d\psi^2,
\end{eqnarray}
where the new form of metric function $\tilde{f}^{\lbrace1\rbrace}$ is given as follows
\begin{eqnarray}\label{metricf1}
\tilde{f}^{\lbrace1\rbrace} = 1-\frac{\tilde{m}(r,\theta)}{|r|^2} 
= 1-\frac{\tilde{m}(r,\theta)}{r^2+a^2\cos^2\theta}.
\end{eqnarray}
Since the function $f(r)$ has to be transformed twice under the complex transformation, 
we need to keep track of the order of the transformations. Now we employ the following 
transformations in order to transform the metric (\ref{metric1}) into the 
$(r,\psi)$-plane
\begin{eqnarray}\label{complex2}
u &=& u^{\prime} +i b\cos\chi_2,\quad r=r^{\prime}-i b\cos\chi_2,\nonumber\\
id\chi_2 &=& -\cos\chi_2 d\psi,\quad \mbox{with}\quad \chi_2=\theta,\nonumber\\
du &=& du^{\prime}-b\cos^2\theta d\psi, \quad dr=dr^{\prime}+b\cos^2\theta d\psi.
\end{eqnarray}
When these transformations are applied directly to the metric (\ref{metric1}), yields
\begin{eqnarray}\label{metric3}
ds^2 &=& -du^2-2dudr+\left(1-\tilde{f}^{\lbrace2\rbrace}\right)\left(du-a\sin^2\theta 
d\psi\right)^2+2a\sin^2\theta dR d\phi+\rho^2 d\theta^2 \nonumber\\&&
+\left(R^2+a^2\right)\sin^2\theta d\phi^2+r^2\cos^2\theta d\psi^2.
\end{eqnarray}
Keep in mind that once again we need to choose the function $R(r)=Re(r)$ to protect 
the geometry of the first plane under these complex transformations. On imposing 
$R=r^{\prime}$ and omitting the prime, we arrive on the new form of the metric 
containing two distinct spin parameters
\begin{eqnarray}\label{metric4}
ds^2 &=& -du^2-2dudr+\left(1-\tilde{f}^{\lbrace 1,2\rbrace}\right)
\left(du-a\sin^2\theta d\phi -b\cos^2\theta d\psi \right)^2 
+2a\sin^2\theta dr d\phi 
\nonumber\\&&
+2b\cos^2\theta dr d\psi +\rho^2 d\theta^2+\left(r^2+a^2\right)\sin^2\theta 
d\phi^2 +\left(r^2+b^2\right)\cos^2\theta d\psi^2
\end{eqnarray}
with
\begin{eqnarray}\label{rho}
\rho^2=r^2+a^2\cos^2\theta+b^2\sin^2\theta.
\end{eqnarray}
The metric function $\tilde{f}^{\lbrace 1,2\rbrace}$ is complexified as follows
\begin{eqnarray}\label{metrcf2}
\tilde{f}^{\lbrace 1,2\rbrace}=1-\frac{\tilde{m}(|r|,\theta)}{|r|^2
+a^2\cos^2\theta}
=1-\frac{\tilde{m}_{\tilde{\alpha},\tilde{\beta}}(r^{\prime},\theta)}
{r{^{\prime}}^2+a^2\cos^2\theta+b^2\sin^2\theta}
=1-\frac{\tilde{m}_{\tilde{\alpha},\tilde{\beta}}(r,\theta)}{\rho^2}.
\end{eqnarray}
On the other hand, the mass function 
$\tilde{m}_{\tilde{\alpha},\tilde{\beta}}(r,\theta)$ is complexified
\begin{eqnarray}\label{mass1}
m(r) = \tilde{m}(|r|,\theta) = 
\tilde{m}_{\tilde{\alpha},\tilde{\beta}}(r^{\prime},\theta)
= \tilde{m}_{\tilde{\alpha},\tilde{\beta}}(r,\theta)
= \mu\Bigg[\frac{r^{3+\tilde{\alpha}}(\rho^2)^{-\tilde{\alpha}}}
{r^{3+\tilde{\alpha}}(\rho^2)^{-\tilde{\alpha}/2}
+q_e^{3}r^{\tilde{\beta}}(\rho^2)^{-\tilde{\beta}/2}}\Bigg]^{4/3},
\end{eqnarray}
where $\tilde{\alpha}$ and $\tilde{\beta}$ are two real numbers appearing during 
the complexification. For the matter of simplicity, we set 
$\tilde{\alpha}=\tilde{\beta}=0$, in this case the mass function looks like
\begin{equation}\label{massf}
    m(r) = \mu \left(\frac{r^3}{r^3 + q_e^3}\right)^{4/3}.
\end{equation}
Now let us transform the metric (\ref{metric4}) into Boyer-Lindquist 
coordinates with the help of the following coordinate transformations
\begin{eqnarray}\label{bltr}
du = dt-g_{1}(r)dr,\quad d\phi=d\phi^{\prime}-g_{2}(r)dr,\quad d\psi=d\psi^{\prime}-g_{3}(r)dr.
\end{eqnarray}
On imposing the conditions $g_{tr}=g_{r\phi\prime}=g_{r\psi\prime}=0$, in 
order to solve (\ref{bltr}) for $g_{1}(r)$, $g_{2}(r)$, and $g_{3}(r)$, we 
obtain
\begin{eqnarray}\label{gs}
g_{1}(r)=\frac{\Pi}{\Delta},\quad g_{2}(r)=\frac{\Pi}{\Delta}\frac{a}{r^2+a^2},
\quad g_{3}(r)=\frac{\Pi}{\Delta}\frac{b}{r^2+b^2},
\end{eqnarray}
where $\Pi$ and $\Delta$ are given by
\begin{eqnarray}
\label{piex}
\Pi=\left(r^2+a^2\right)\left(r^2+b^2\right),\quad 
\Delta=\left(r^2+a^2\right)\left(r^2+b^2\right)-m(r)r^2.
\end{eqnarray}
We further apply the Boyer-Lindquist coordinate transformations (\ref{bltr}) on 
the metric (\ref{metric4}), as a consequence we get the final form of the 
five-dimensional rotating spacetime
\begin{eqnarray}\label{metric5}
ds^2&=&-dt^2+\left(1-\tilde{f}^{\lbrace 1,2\rbrace}\right)
\left(dt-a\sin^2\theta d\phi -b\cos^2\theta d\psi\right)^2
+\frac{r^2\rho^2}{\Delta}dr^2 \nonumber\\&&
+\rho^2 d\theta^2+\left(r^2+a^2\right)\sin^2\theta d\phi^2 
+\left(r^2+b^2\right)\cos^2\theta d\psi^2.
\end{eqnarray}
Here the angles $\phi$ and $\psi$ lies in the interval $[0,2\pi]$ while the angle 
$\theta$ takes values from the interval $[0,\pi/2]$. When we substitute function 
$\tilde{f}^{\lbrace 1,2\rbrace}$ into (\ref{metric5}), eventually it reads simply
\begin{eqnarray}\label{metricf}
ds^2&=&-dt^2 +\frac{m(r)}{\rho^2} \left(dt-\omega \right)^2 
+\frac{r^2 \rho^2}{\Delta}dr^2 +\rho^2 d\theta^2 \nonumber\\
&&  +\left(r^2 +a^2\right)\sin^2\theta d\phi^2 
+\left(r^2 +b^2\right)\cos^2\theta d\psi^2,
\end{eqnarray}
where $\omega$ is defined as follows
\begin{equation}
    \omega = a\sin^2\theta d\phi + b\cos^2\theta d\psi.
\end{equation}
We must emphasize that as like the Meyers-Perry black hole, the rotating 
five-dimensional electrically charged Bardeen spacetime has two rotation parameters 
($a$ and $b$). Note that the rotation is along the ($\phi, \psi$) axes, where $0\leq 
\phi\leq 2\pi$ and $0\leq\psi\leq\pi/2$. We recover the Meyers-Perry spacetime, in 
the limit $q_e\to 0$. Hence, we reach at the conclusion that the presence of charge 
$q_e$ provides deviation from the Meyers-Perry spacetime.  
 
Now we are going to check the validity of the rotating five-dimensional spacetime 
by computing the source nonlinear electrodynamics expressions. The gauge potential 
(\ref{gauge}) for the rotating spacetime gets modify \cite{Aliev:2004ec} as follows
\begin{equation}\label{vecp}
    A^{\mu}= -\frac{\mu r^7}{q_e \left(r^3 + q_e^3\right)^{7/3}}
 \left(\delta^{\mu}_t  -a \sin^2 \theta \delta^{\mu}_{\phi} 
 -b \cos^2 \theta \delta^{\mu}_{\psi} \right).
\end{equation}
On using this gauge potential, it is easy to compute the field strength $(\mathcal{F})$ 
in case of the rotating spacetime which turns out to be in following form
\begin{equation}\label{fs}
    \mathcal{F} = \frac{\mu^2 r^{12} \left[-G m r^2 + \rho^2
    \lbrace J (a^2-b^2) \cos^2 \theta + N (r^2 +a^2) \rbrace \right]}
    {2 \Delta \rho^4 q_e^2 \left(r^3 + q_e^3\right)^{20/3}},
\end{equation}
where $G= 49 q_e^6 (r^4+b^4) -4 b^2 r^5 (r^3+2q_e^3)  
+2 (a^2-b^2) \cos^2 \theta (49b^2 q_e^6 +47 q_e^6 r^2 -4q_e^3 r^5 -2r^8)
 +49 q_e^6 (a^2-b^2) \cos^4 \theta +94 b^2 q_e^6 r^2$, 
 $J= 49 q_e^6 r^2 (a^2+b^2) -4r^7 (r^3+2q_e^3) +49 a^2 b^2 q_e^6 +45 q_e^6 r^4$, 
and $N= 49q_e^6 (r^2+b^2)^2 -4 b^2 r^2 (r^3+q_e^3)^2 $.
We immediately recover $\mathcal{F}$ exactly as (\ref{nfs}) when substitute $a=b=0$ 
in (\ref{fs}). Our next task is to determine the source for the rotating spacetime. 
We solve Einstein field equations of the rotating five-dimensional spacetime in 
order to calculate $\mathcal{L}$ and $\mathcal{L}_\mathcal{F}$. It turns out that the 
Lagrangian density is given by
\begin{eqnarray}\label{rl}
  \mathcal{L}(r) = \frac{49 \Delta m^{\prime \prime} \rho^4 q_e^6 
  (\Delta +mr^2) +2 r m^{\prime} (a^2-b^2)\cos^2 \theta
  \lbrace S -m T r^2 +49 q_e^6 (\Delta +mr^2)^2\rbrace}
  {\rho^2 \lbrace U (a^2-b^2)\cos^2 \theta 
  +49 \Delta q_e^6 (r^2+a^2)(r^2+b^2)^2 \rbrace},
\end{eqnarray}
where $S= (a^2-b^2)\cos^2 \theta \lbrace 49 q_e^6 (r^2+a^2)^2 (r^2+b^2)
-mr^2 (2 r^8 +49 a^2 q_e^6 +51 q_e^6 r^2 +4 q_e^3 r^5) \rbrace $, 
$T= 6r^{10} +2b^2 r^8 +12 q_e^3 r^7 +4b^2q_e^3r^5 +55 q_e^6 r^4 
+49 a^2 b^2 q_e^6 +q_e^6 (49 a^2 + 51b^2) +49 q_e^6 (\Delta +mr^2)^2$, and 
$U = 49 q_e^6 (\Delta +mr^2)^2  -mr^2 \lbrace 4r^{10} 
+8q_e^3 r^7 +53 q_e^6 r^4 +49 a^2 b^2 q_e^6 +49 q_e^6 r^2 (a^2+b^2) \rbrace$. 
On the other hand, $\mathcal{L}_\mathcal{F}$ can be expressed as follows
\begin{eqnarray}\label{rlf}
  \mathcal{L}_\mathcal{F}(r) = -\frac{ \Delta q_e^2 (r^2+a^2)(r^3+q_e^3)^{20/3}
  \lbrace r m^{\prime \prime} \rho^2(r^2+b^2)-m^{\prime}V \rbrace}
  {4 \mu^2 \rho^2 r^{13} \lbrace U (a^2-b^2)\cos^2 \theta 
  +49 \Delta q_e^6 (r^2+a^2)(r^2+b^2)^2\rbrace},
\end{eqnarray}
where $V= 3r^4 +b^2 (4r^2 +b^2) - (r^2-b^2) (a^2-b^2) \cos^2 \theta$. 
We emphasize that the Einstein field equations for the rotating five-dimensional 
Bardeen spacetime have cumbersome forms that's why we are not going to show them in 
the paper. The source equations \eqref{rl} and \eqref{rlf} have been obtained by using 
the field equations with particular choice of the vector potential given in 
\eqref{vecp}. When these equations substitute back into the field equations, they 
ultimately satisfy them.

We further substitute both the rotation parameter $a$ and $b$ equal to zero into 
(\ref{rl}) and (\ref{rlf}), yields
\begin{equation}
\mathcal{L}(r) = \frac{ \mu q_e^3 \left(3q_e^3 -4r^3\right)}
{\left(r^3 +q_e^3\right)^{10/3}}, \quad 
\mathcal{L}_{\mathcal{F}}(r) = 
 \frac{\left(r^3 +q_e^3\right)^{10/3}}{7 \mu q_e r^9},
\end{equation}
which are exactly similar to that of the static spacetime. In other words, 
we recover the nonrotating source expressions when both rotation parameters are set 
equal to zero. It is noticeable that the spacetime (\ref{metricf}) does not have any 
dependence upon the $t$, $\phi$, and $\psi$ coordinates which leads to the three 
Killing vectors, namely, $\partial/\partial{t}$, $\partial/\partial{\phi}$, and 
$\partial/\partial{\psi}$. They correspond to the conserved quantities of physical 
interest: the energy and the two angular momenta in the $r-\phi$ plane and the 
$r-\psi$ plane.

\section{Energy conditions}
\label{enc}
In order to discuss the energy conditions and the matter associated with the 
spacetime (\ref{metricf}), we need to compute the orthonormal basis by 
considering the locally nonrotating frames (LNRF). The dual basis vectors 
carried by the local observer \cite{Aliev:2004ec} take the following form 
\begin{eqnarray}
e_{(t)} &=& \left|g_{tt} +\Omega_{\phi}g_{t \phi} 
+\Omega_{\psi} g_{t \psi}\right|^{-1/2} 
\left(\frac{\partial}{\partial t} +\Omega_{\phi} \frac{\partial}{\partial \phi} 
+\Omega_{\psi} \frac{\partial}{\partial \psi}\right), \nonumber\\
e_{(r)} &=& \frac{1}{\sqrt{|g_{r r}|}} \frac{\partial}{\partial r}, \nonumber\\
e_{(\theta)} &=& \frac{1}{\sqrt{g_{\theta \theta}}} \frac{\partial}
{\partial \theta}, \nonumber\\
e_{(\phi)} &=& \left(\frac{g_{\phi \phi} g_{\psi \psi}-g_{\phi \psi}^2}
{g_{\psi \psi}}\right)^{-1/2} 
\left(\frac{\partial}{\partial \phi} - \frac{g_{\phi \psi}}
{g_{\psi \psi}} \frac{\partial}{\partial \psi}\right), \nonumber\\
e_{(\psi)} &=& \frac{1}{\sqrt{g_{\psi \psi}}} \frac{\partial}{\partial \psi}, 
\end{eqnarray}
where the angular velocities of the spacetime corresponding to the $\phi$-axis and 
$\psi$-axis are defined \cite{Aliev:2004ec} by
\begin{equation}
\Omega_{\phi} = \frac{g_{t \psi} g_{\phi \psi}-g_{t \phi} g_{\psi \psi}}
{g_{\phi \phi} g_{\psi \psi}-g_{\phi \psi}^2}, \quad
\Omega_{\psi} = \frac{g_{t \phi} g_{\phi \psi}-g_{t \psi} g_{\phi \phi}}
{g_{\phi \phi} g_{\psi \psi}-g_{\phi \psi}^2}.
\end{equation}
The corresponding one-forms of the dual basis can be written in the following matrix 
form:
\begin{equation}\label{orthob}
e^{(a)}_{\mu}=\left(\begin{array}{ccccc}
\sqrt{ \mp \left(g_{tt} +\Omega_{\phi} g_{t \phi}
+\Omega_{\psi} g_{t \psi} \right)} & 0 & 0 & 0 & 0\\
0 & \sqrt{\pm g_{rr}} & 0 & 0 & 0\\
0 & 0 & \sqrt{g_{\theta \theta}} & 0 & 0\\
-\Omega_{\phi} \sqrt{g_{\phi \phi} g_{\psi \psi}
-g_{\phi \psi}^2}/\sqrt{g_{\psi \psi}} & 0 & 0 
& \sqrt{g_{\phi \phi} g_{\psi \psi}
-g_{\phi \psi}^2}/\sqrt{g_{\psi \psi}} & 0\\
-\Omega_{\phi} g_{\phi \psi}/\sqrt{g_{\psi\psi}} 
-\Omega_{\psi}\sqrt{g_{\psi \psi}} & 0 & 0 
& {g_{\phi\psi}}/{\sqrt{g_{\psi\psi}}} & \sqrt{g_{\psi\psi}} 
\end{array}\right).
\end{equation}
\begin{figure*}
\includegraphics[scale=0.53]{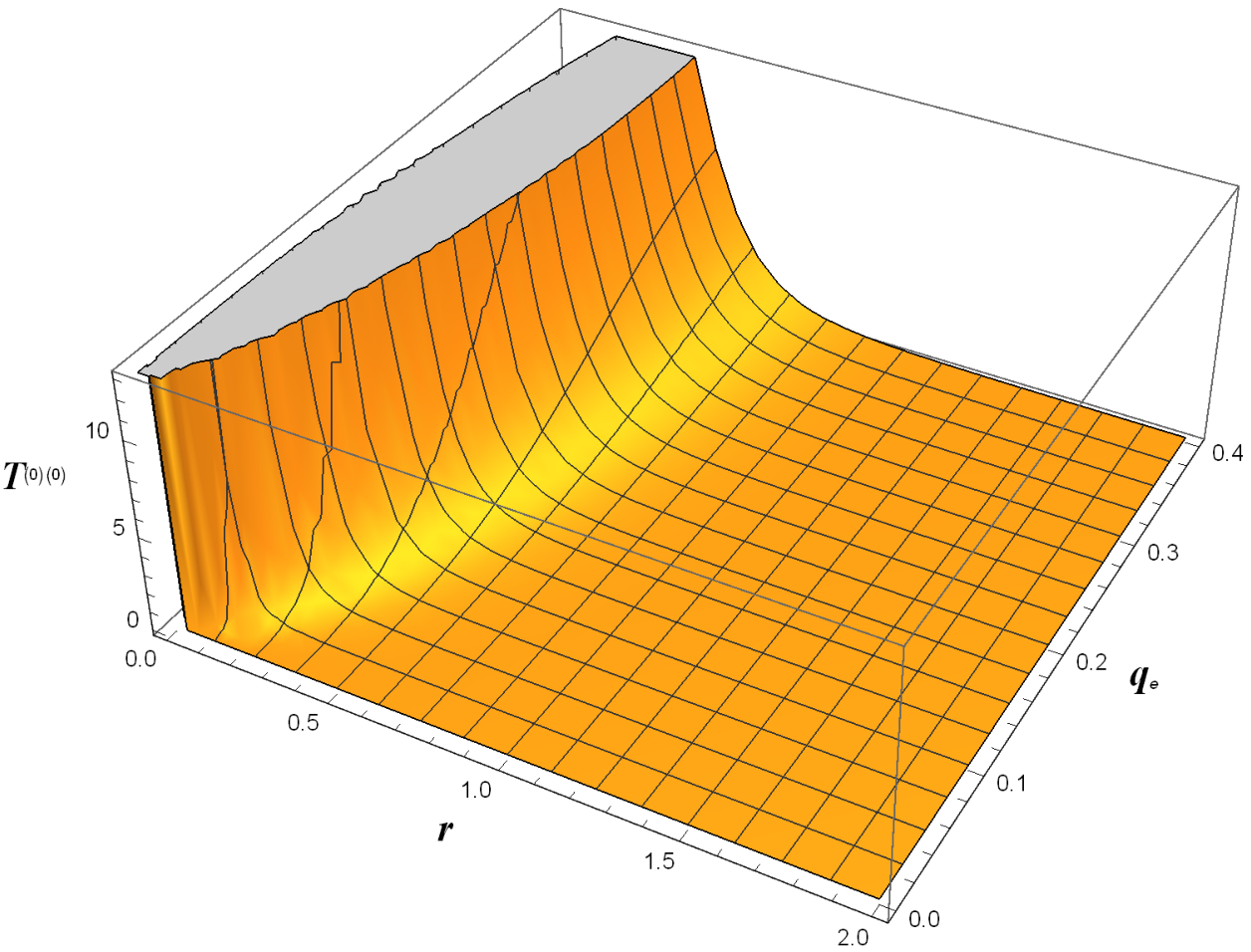}
\includegraphics[scale=0.53]{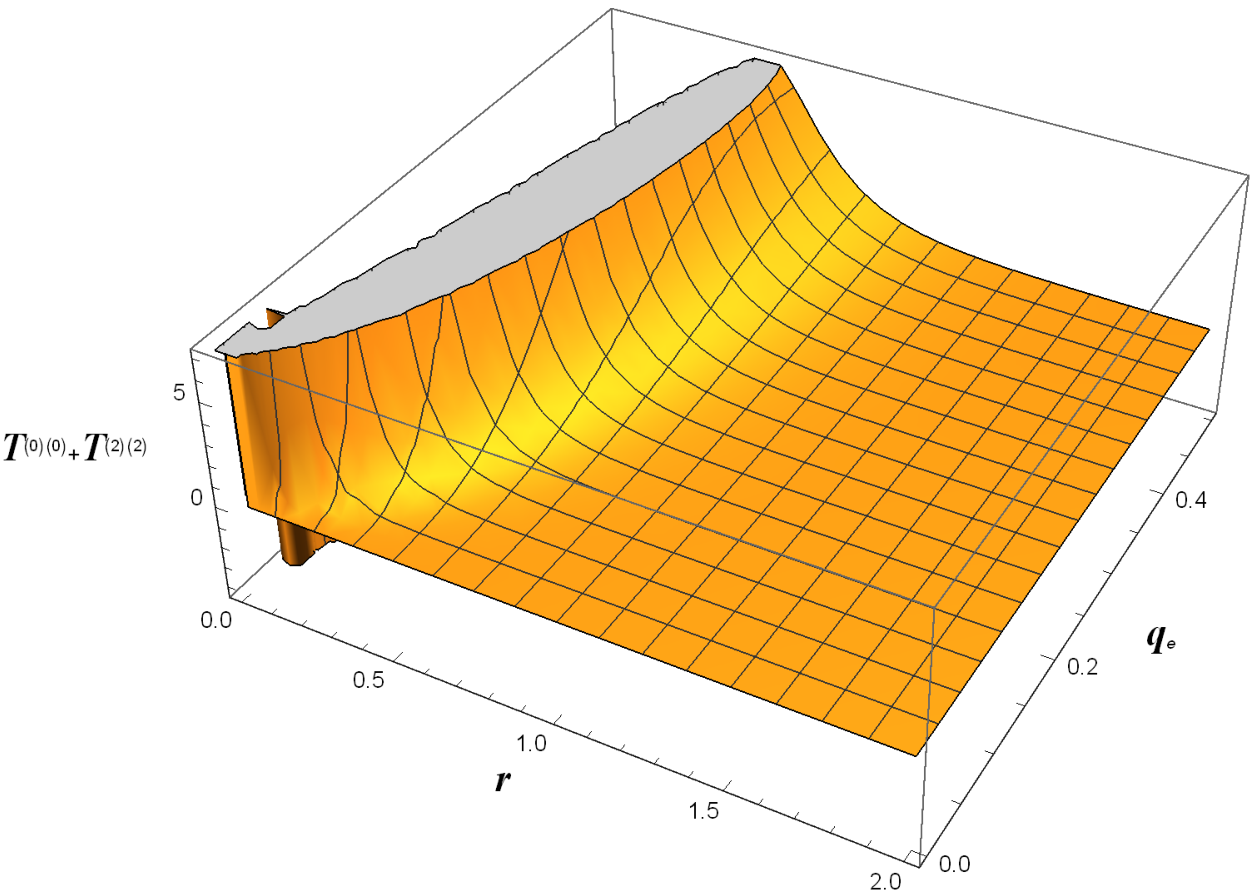}\\
\includegraphics[scale=0.53]{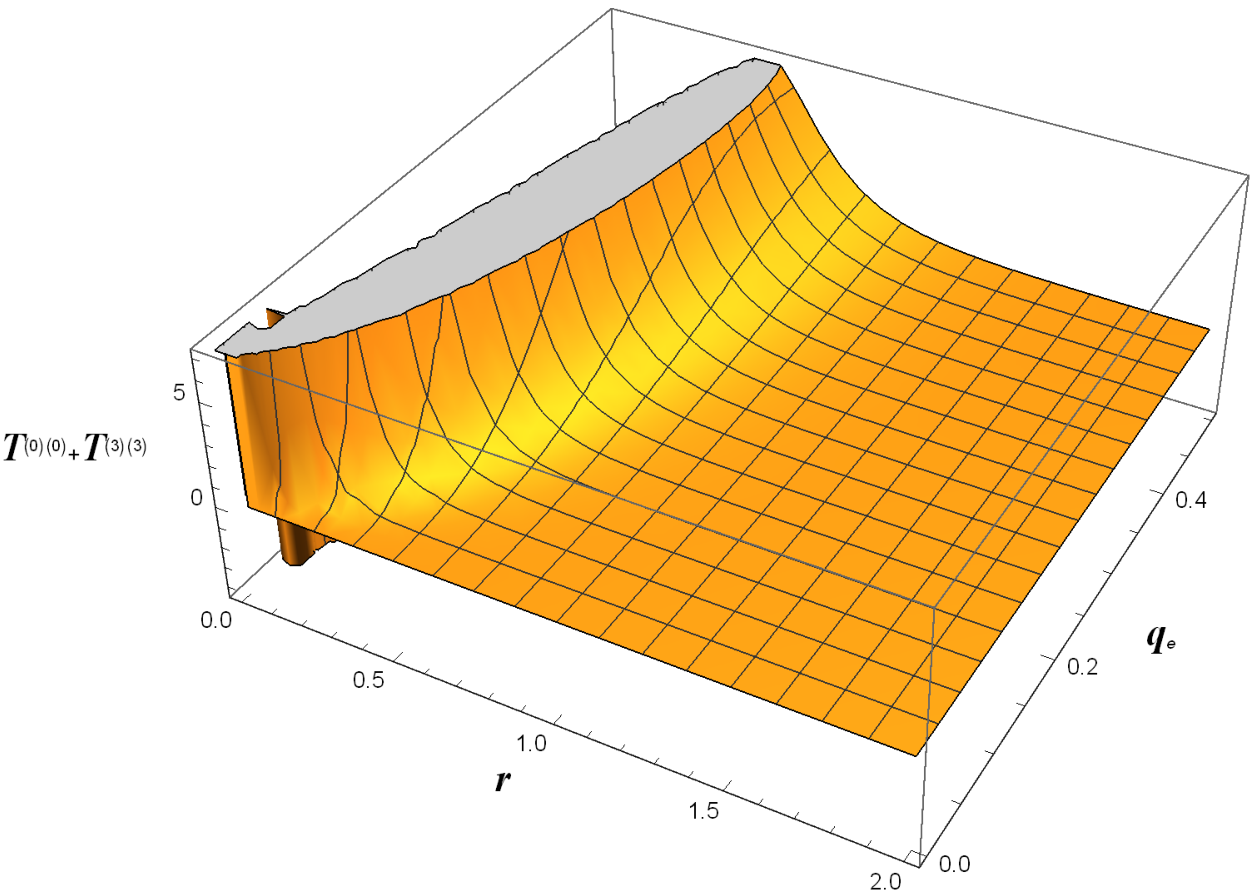}
\includegraphics[scale=0.53]{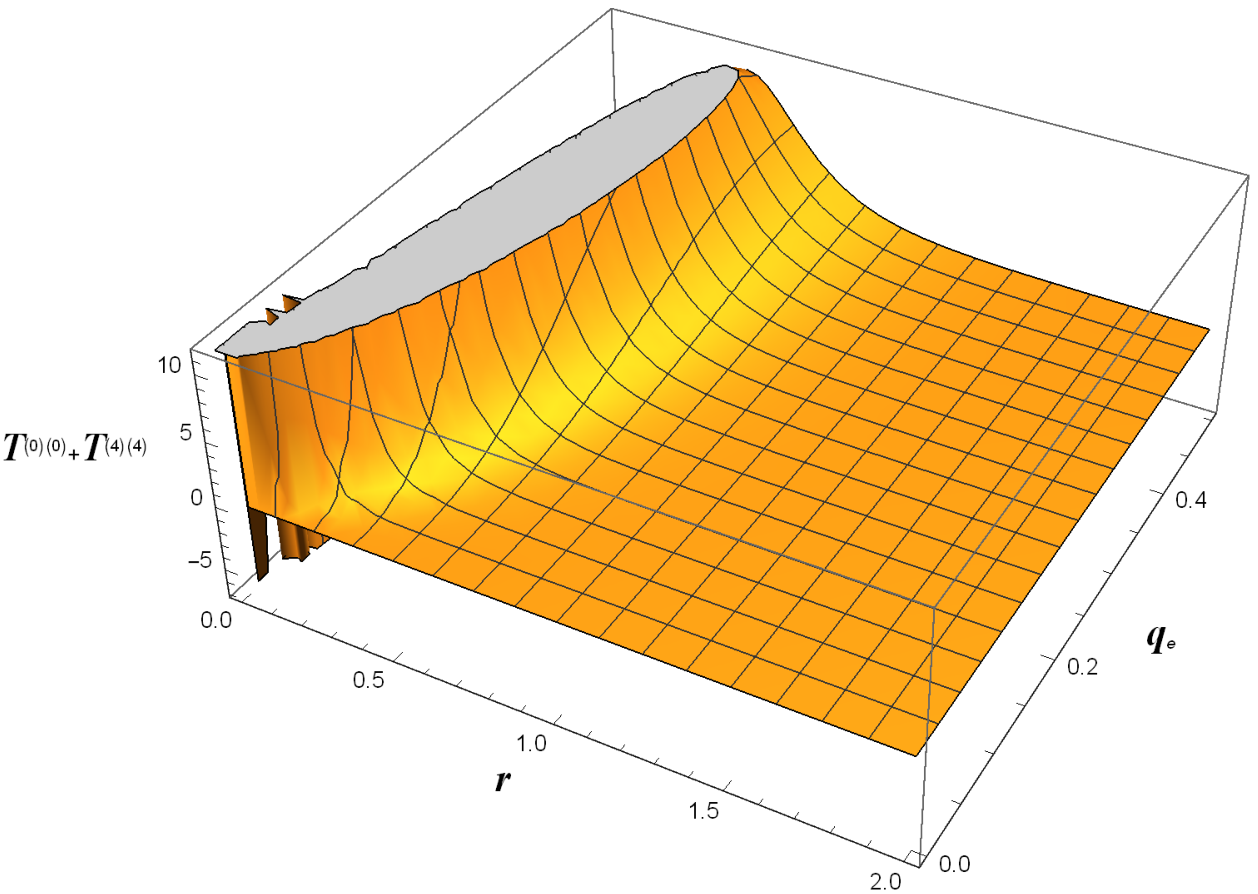}
\caption{\label{energ} (color online). Plot showing the behavior of the 
energy conditions of the rotating spacetime for the set of values 
$a=0.4, b=0, \mu=1$.}
\end{figure*}
Here the signatures reflect that the considered region whether it is outside the event 
horizon or inside the Cauchy horizon correspond to the choice $(-,+)$, respectively. 
The energy-momentum tensor can be expressed in terms of orthonormal basis which reads 
\begin{equation}
    T^{(a) (b)} = e_{\mu}^{(a)} e_{\mu}^{(b)} G^{\mu \nu},
\end{equation}
which turned to be in a diagonal form. The computation reveals that the components of 
$T^{(a) (b)}$ are rather cumbersome. Therefore, we consider $\theta =0,\pi$ for the
matter of simplicity. The nonzero components of $T^{(a) (b)}$ for $\theta =0,\pi$ 
are given as follows
\begin{eqnarray}\label{emtc}
  T^{(0)(0)} &=& \frac{m^{\prime} (3r^2+a^2)(r^2+b^2)^2 
  -b^2 m^{\prime \prime} r \Delta }
  {2 r^3 (r^2 +a^2)[b^2 m +(r^2 +a^2)(r^2 +b^2)]}, \nonumber\\
  T^{(1)(1)} &=& -\frac{m^{\prime} (3r^2 +a^2)}{2 r (r^2 +a^2)^2 }, \nonumber\\
  T^{(2)(2)} &=& -\frac{m^{\prime \prime}r (r^2 +a^2) + 2m^{\prime} a^2}
   {2 r (r^2 +a^2)^2}, \nonumber\\
  T^{(3)(3)} &=& -\frac{m^{\prime \prime}r (r^2 +a^2) + 2 m^{\prime} a^2}
   {2 r (r^2 +a^2)^2},\nonumber\\
  T^{(4)(4)} &=& -\frac{m^{\prime \prime} (r^2 +a^2)^2 (r^2 +b^2)^2 
  - m^{\prime} b^2 (3r^2 +a^2)\Delta}
   {2 r^3 (r^2 +a^2)^2 [b^2 m+(r^2 +a^2)(r^2 +b^2)]}.
\end{eqnarray}
As can be seen from the components of energy-momentum tensor that for both 
$T^{(0)(0)}$ and $T^{(4)(4)}$ components, there is a 
dependence on both angular momenta. If 
we set $b=0$ in (\ref{emtc}), it turns out that $T^{(0)(0)} = -T^{(1)(1)}$. 
The typical behavior of the energy conditions can be seen from the Fig~\ref{energ}. 
We see that there is a violation of weak energy condition in $T^{(0)(0)}+T^{(4)(4)}$, 
but only by a small amount (cf. Fig~\ref{energ}). It is the black hole rotation 
which leads to the violation of energy condition for any physically reasonable mass. 
The violation of weak energy condition cannot be prevented in the rotating regular 
black holes \cite{Bambi:2013ufa,Neves:2014aba}. Now we are going to discuss some 
important properties of the rotating spacetime in next section.

\section{Nature of rotating spacetimes}
\label{prop}
In this section, we discuss some important properties of the rotating 
five-dimensional electrically charged Bardeen regular spacetime comprehensibly.

\subsection{Curvature invariants}
\label{curvin}
\begin{figure*}
\includegraphics[scale=0.75]{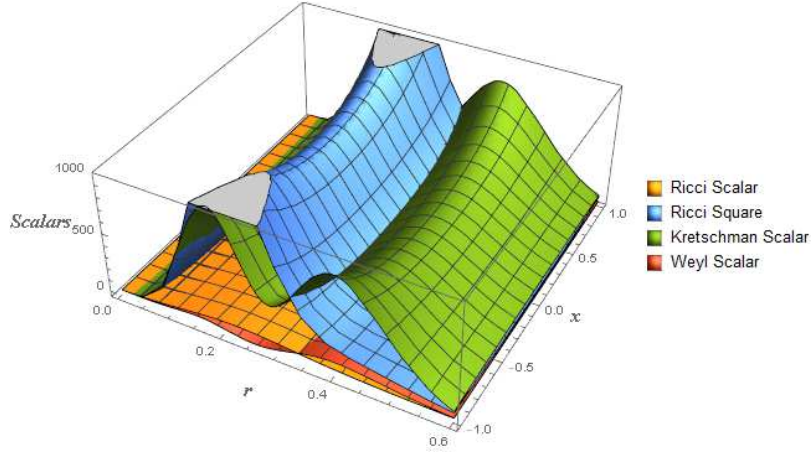}
\caption{\label{curv} (color online). Plot showing the behavior of 
curvature invariants of the rotating spacetime for the set of values $a=0.5, b=0.6, q_e=0.4, \mu=1$ ($x=\cos \theta$).}
\end{figure*}

We calculate the curvature invariants of the rotating five-dimensional 
electrically charged Bardeen regular spacetime (\ref{metricf}) to verify the regularity 
of them. In order to do so we compute the expressions of the curvature 
scalars and find their limiting values. At first we compute the Ricci scalar which takes the form 
\begin{eqnarray}
R &=& g^{\mu \nu}R_{\mu \nu} = R^{\mu}_{\mu} =\frac{m'' r +2 m'}{r \rho^2},
\end{eqnarray}
where $m$ is the mass function defined in (\ref{massf}) and prime ($'$) denotes 
the derivative with respect to the radial coordinate $r$. The other curvature 
scalar which is contraction of the Ricci tensor is given by
\begin{eqnarray}
\mathcal{R} &=& R_{\mu \nu}R^{\mu \nu} 
=\frac{\rho^4 m''^2 r^2  +2 m' m'' \rho^2 (\rho^2 -2r^2) r 
+ m'^2 (3\rho^4 -4 \rho^2 r^2 +7r^4)}{2 r^2 \rho^8}.
\end{eqnarray}
Another important curvature scalar known as the Kretschmann scalar 
$(K= R_{\mu \nu \gamma \delta} R^{\mu \nu \gamma \delta})$ can be computed as follows
\begin{eqnarray}
K &=& \frac{\left[\rho^8 m''^2 -4 \rho^4 \left(2 \rho^2 m' r + \gamma m \right) m'' 
+24 \gamma \alpha m^2 \right]r^2 -8 \rho^2 \xi m m' r 
+2 \rho^4 \delta m'^2}{r^2 \rho^{12}},
\end{eqnarray}
where $\gamma= \rho^2 -4r^2$, $\delta= \rho^4 -6 \rho^2 r^2 +16r^4$, and 
$\xi= \rho^4 -16 \rho^2 r^2 + 24 r^4$. We further compute the Weyl invariant 
($W1R = C_{\mu \nu \gamma \delta} C^{\mu \nu \gamma \delta}/8$) for the spacetime  
(\ref{metricf}), which is given by
\begin{eqnarray}
W1R &=& \frac{\left(11\rho^8 m''^2 +432 \gamma \eta m^2 
-72\rho^4 \gamma m m'' \right) r^2 
-2 \rho^2 \left(\rho^4 \lambda m'' +72 \xi m \right) m' r 
+ \rho^4 \sigma m'^2}{144 r^2 \rho^{12}}, \nonumber\\
\end{eqnarray}
where $\eta= 3 \rho^2 -4r^2$, $\lambda= 5\rho^2 +54 r^2$, and 
$\sigma =17 \rho^4 -180 \rho^2 r^2 + 504 r^4$. 
It is noticeable that the curvature invariants of the rotating five-dimensional 
electrically charged Bardeen regular black holes have very tedious mathematical 
expressions. Therefore, it is bit difficult to visualize analytically the nature of 
them. For this reason, we plot the curvature invariants and the illustration of these 
invariants can be seen in Fig.~\ref{curv}. This illustration confirms that all 
the curvature invariants are well behaved even at the origin $r=0$. The finiteness of 
scalar invariants at $r=0$ is in support of the existence of the de Sitter belt around 
the origin thereby the spacetime does not terminate in signaling a singular nature.

Apart from the aforementioned curvature invariants, there exists other curvature 
invariants as well. We are also interested in calculation of trace-free 
Ricci invariants which can be expressed by $r_1 =S_{\mu}^{\nu} S_{\nu}^{\mu}/4$, 
$r_2 = -S_{\mu}^{\nu} S_{\nu}^{\gamma} S_{\gamma}^{\mu}/8$, and
$r_3 = S_{\mu}^{\nu} S_{\nu}^{\gamma} S_{\gamma}^{\delta} S_{\delta}^{\mu}/16$. 
The trace-free Ricci tensor is given by 
$S_{\mu \nu} = R_{\mu \nu} - R g_{\mu \nu}/D$, where $D$ represents the number of 
dimensions. Hence, these trace-free Ricci invariants for the rotating spacetime 
(\ref{metricf}) can be expressed in the following forms
\begin{eqnarray}
r_1 &=& \frac{3 {m^{\prime \prime}}^2 \rho^4 r^2  
+2 m^{\prime} m^{\prime \prime} \rho^2 r (\rho^2 -10r^2)
+{m^{\prime}}^2(7\rho^4 -20 \rho^2 r^2 +40 r^4)}{40 \rho^8 r^2}, \nonumber\\
r_2 &=& \frac{3(3 m^{\prime}-m^{\prime \prime} r)
\left[{m^{\prime \prime}}^2 r^2 (\rho^2 -r^2)^2 
+ 2 m^{\prime} m^{\prime \prime} \rho^2 r (2 \rho^2 -5 r^2) 
 -{m^{\prime}}^2 (\rho^4 +10 \rho^2 r^2 -20 r^4) \right]}{800 \rho^{10} r^3}, 
 \nonumber\\
r_3 &=& \frac{1}{16000 \rho^{16} r^4} 
\Big[21 {m^{\prime \prime}}^4 \rho^8 r^4 
+28 m^{\prime} {m^{\prime \prime}}^3 \rho^6 r^3 (\rho^2 -10 r^2) \nonumber\\
&& +6 {m^{\prime}}^2 {m^{\prime \prime}}^2 \rho^4 r^2 (29\rho^4 -100\rho^2 r^2 
+260 r^4)-4 {m^{\prime}}^3 m^{\prime \prime} \rho^2 r 
(17 \rho^6 -660 \rho^2 r^4 +210 \rho^4 r^2 +1000 r^6) \nonumber\\
&& +{m^{\prime}}^4 \lbrace 170 r^8 + r^4 (\rho^2 - r^2) (1566 \rho^2 -2322 r^2) 
+ (\rho^2 - r^2)^3 (181\rho^2 + 23r^2) \rbrace \Big].
\end{eqnarray}
We can easily check the limit, $r \rightarrow 0$ for the trace-free Ricci invariants 
which confirms that they do not diverge. This is the confirmation from these curvature 
invariants that the spacetime (\ref{metricf}) is regular everywhere.

\subsection{Horizon structure}
\label{horz}
\begin{figure}[hb]
\includegraphics[scale=0.48]{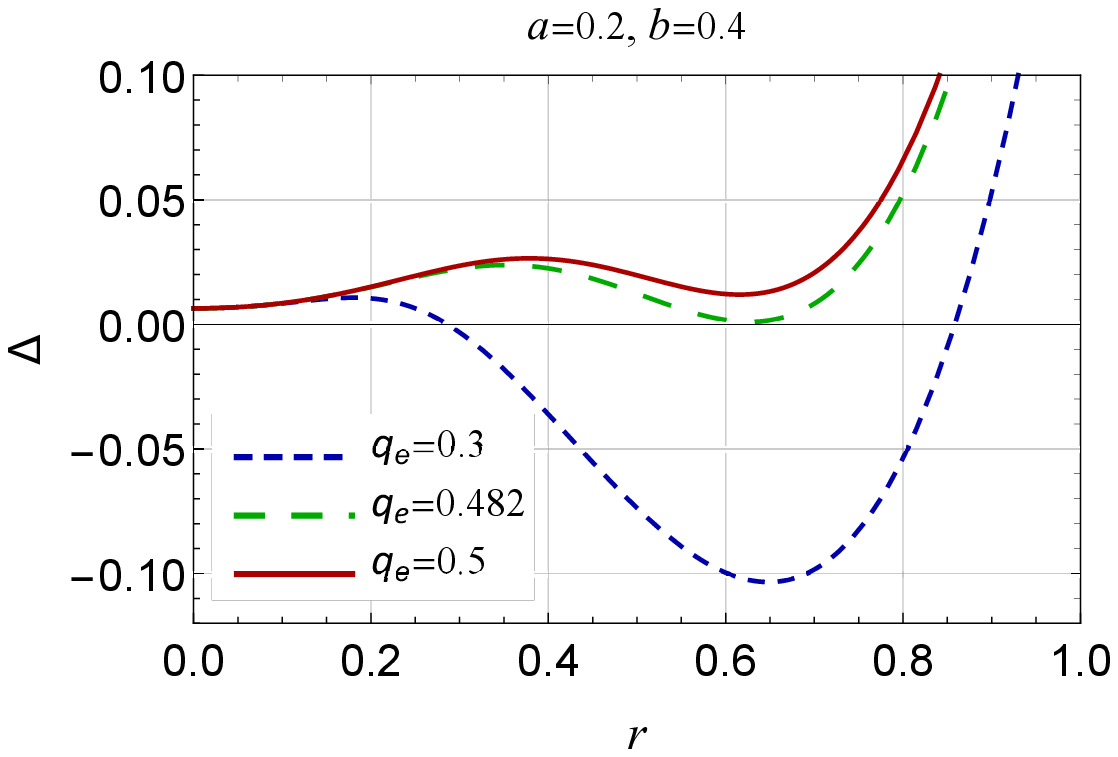}
\includegraphics[scale=0.48]{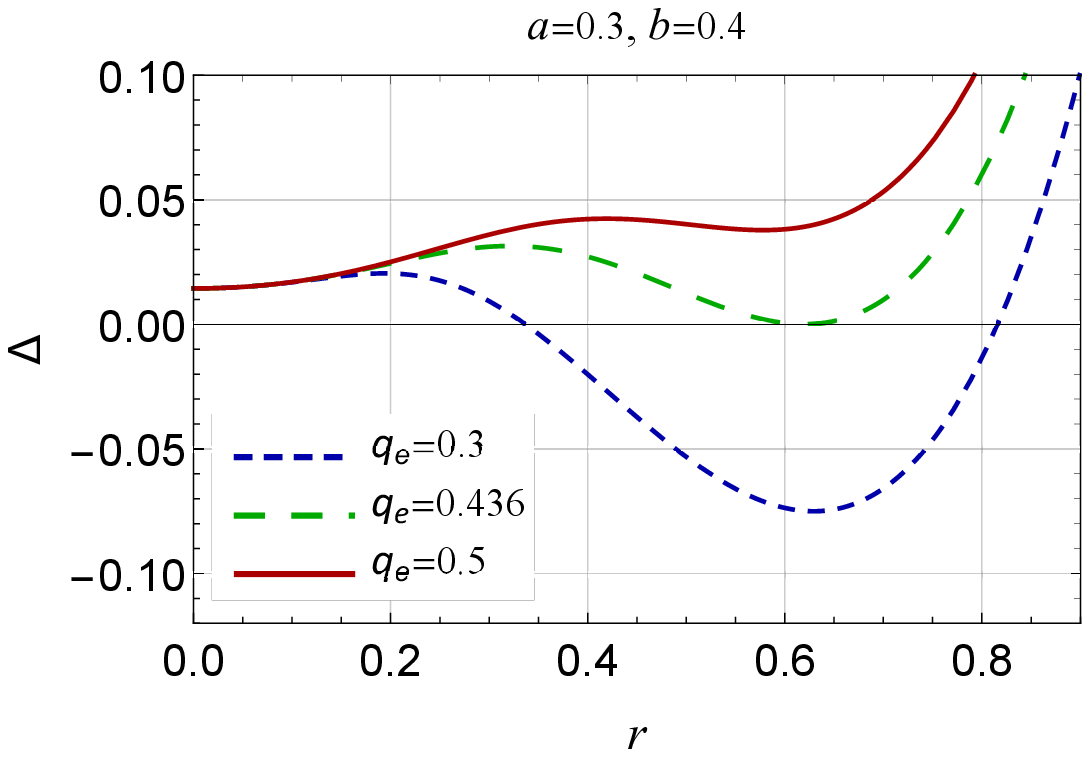}
\includegraphics[scale=0.49]{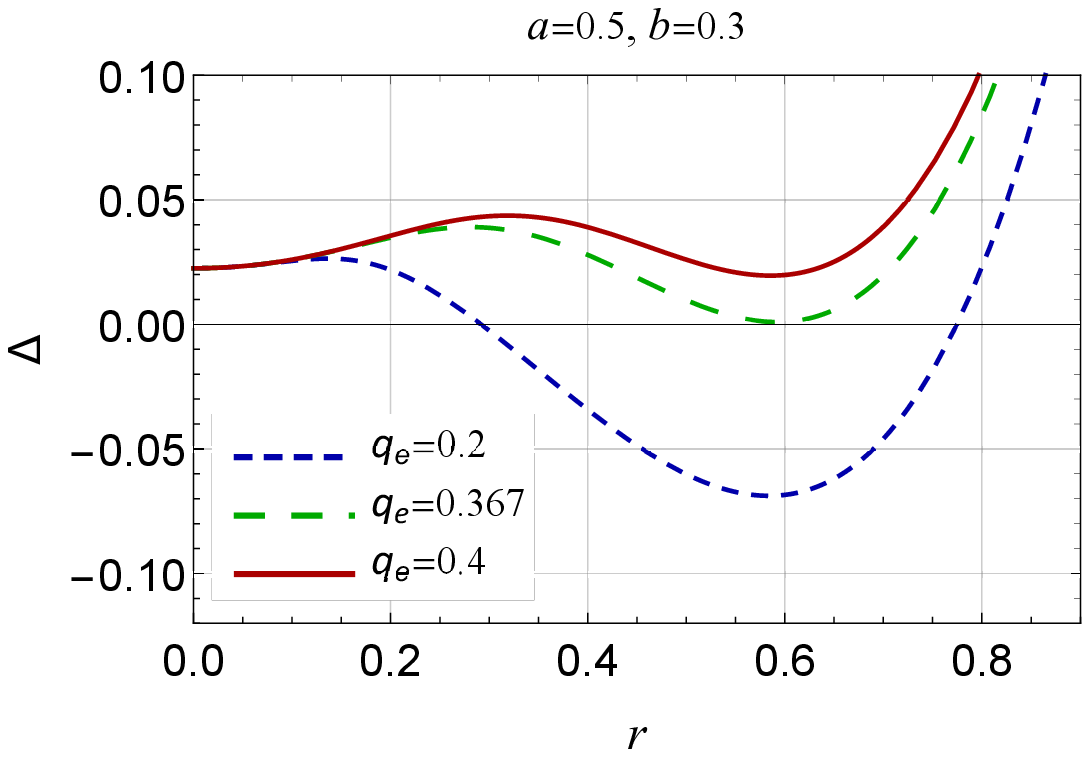}
\caption{\label{horzp} (color online). Plot showing the behavior of horizons of the 
rotating five-dimensional electrically charged Bardeen regular black holes.}
\end{figure}

Next, we discuss the horizons of the rotating five-dimensional electrically charged 
Bardeen regular black hole. It is noticeable that the metric (\ref{metricf}) has a 
singularity at $g^{rr}= \Delta=0$, which corresponds to the horizons of the black 
hole. Keep in mind that this is a coordinate singularity and it can easily be 
removed by a particular choice of transformations. The horizons of the rotating 
five-dimensional electrically charged Bardeen regular black hole are solutions of the 
following equation
\begin{eqnarray}\label{horzeq}
\left(r^2+a^2\right)\left(r^2+b^2\right)-\mu r^2 
\left(\frac{r^3}{r^3 +q_e^3}\right)^{4/3}=0.
\end{eqnarray}
It is clear that the analytical solution of (\ref{horzeq}) is not easily found 
because of the fractional power. In order to understand the nature of (\ref{horzeq}), 
we plot it for different set of values of the parameters $a$, $b$, and $q_e$ 
(cf. Fig.~\ref{horzp}). We find that there exists three different cases in horizons 
profile which correspond to the two horizons (blue dashed line), extremal horizons 
(green dashed line), and no horizon (red line). The two horizons are correspond to 
the event horizon and the Cauchy horizon. The extremal horizons or degenerate 
horizons exist for the precise values of charge $q_e$ (cf. Fig.~\ref{horzp}).

\begin{figure}
\includegraphics[scale=0.49]{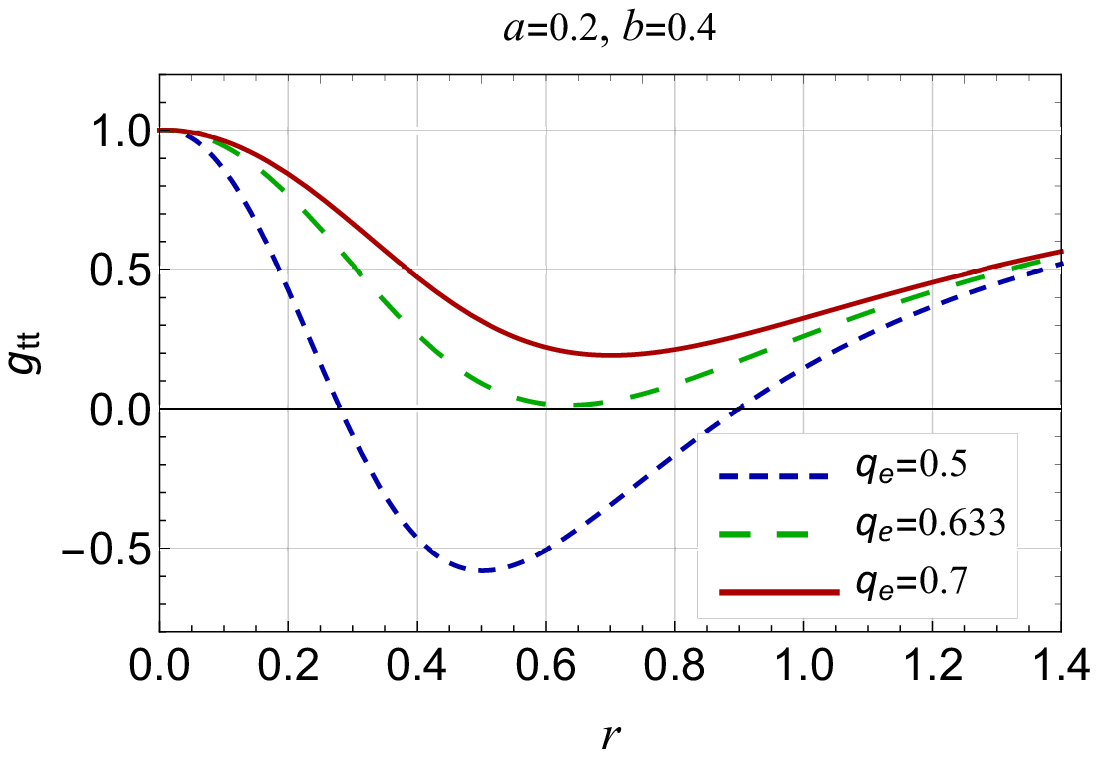}
\includegraphics[scale=0.49]{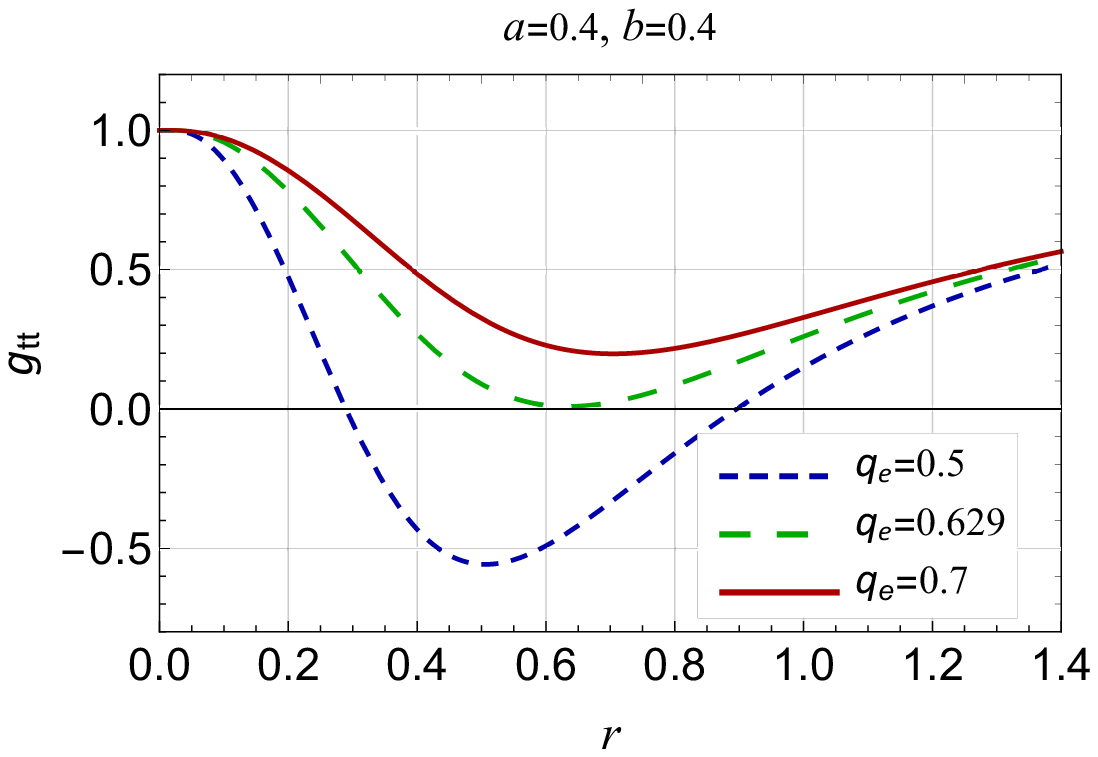}
\includegraphics[scale=0.49]{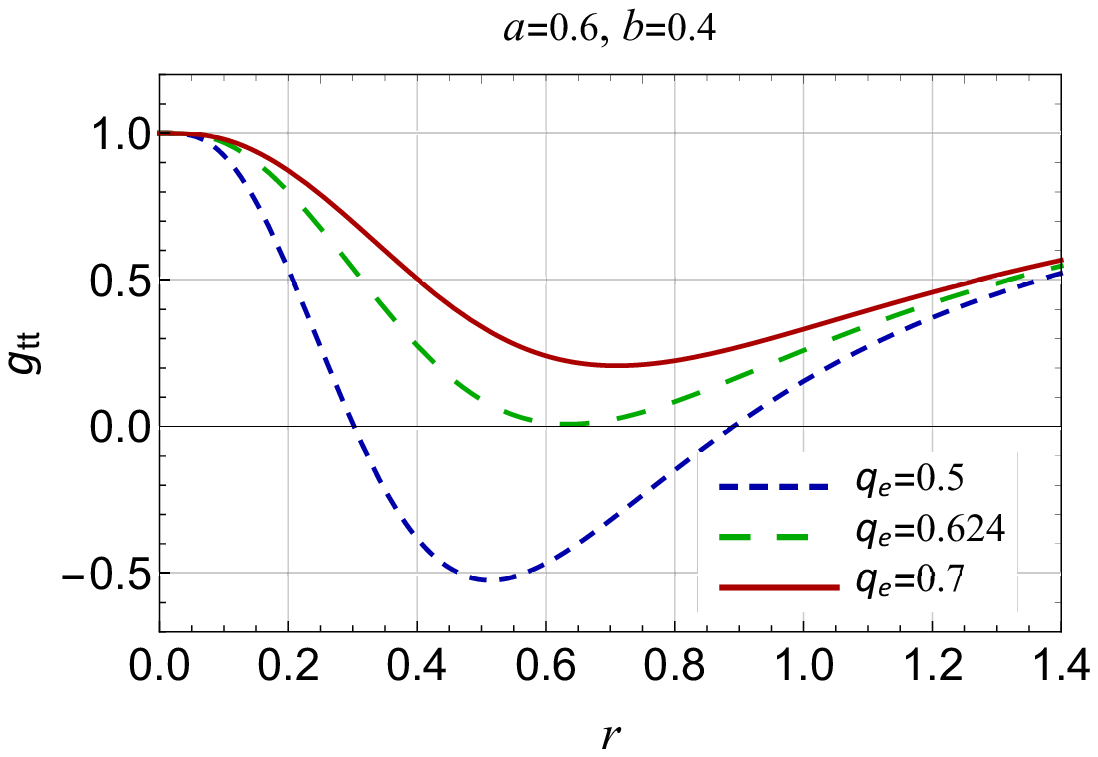}
\caption{\label{horzps} (color online). Plot showing the behavior of 
static limit surface of rotating five-dimensional electrically charged 
Bardeen regular black hole.}
\end{figure}

The static limit surface or infinite redshift surface of the rotating five-dimensional 
electrically charged Bardeen regular black hole can be evaluated by equating the 
coefficient of $dt^2$ to zero, i.e.,
\begin{eqnarray}
r^2+a^2\cos^2\theta+b^2\sin^2\theta - \mu \left(\frac{r^3}{r^3+q_e^3}\right)^{4/3}=0
\end{eqnarray}
depending on the parameters $a$, $b$, and $q_e$ as well as on the angle $\theta$. 
We show the effect of these parameters on the shape of the static limit surface in 
Fig.~\ref{horzps}, by considering different sets of values of $a$, $b$, and $q_e$.
It is found that there exists three different cases which correspond to 
the two roots (blue dashed line), degenerate roots (green dashed line), 
and no root (red line). We illustrate different cases where the static 
limit surfaces have degenerate roots as can be seen in in 
Fig.~\ref{horzps}.

\subsection{Ergoregion}
\label{ergo}
An ergoregion is a spacetime region where timelike Killing vector becomes spacelike. 
In black hole physics, there exists a very special phenomenon according to which a 
Killing field may be timelike in particular regions and spacelike in other regions 
of spacetime. Basically, this region lies outside the event horizon of the rotating 
black holes, i.e., $r_{sls}<r<r_{+}$, where $r_{sls}$ and $r_{+}$ corresponding to 
radius of the static limit surface and radius of the event horizon, respectively. 
On the other hand, the boundary of the ergoregion is known as the ergosphere. An 
ergosphere is a surface of infinite redshift because the norm of the time-translation 
Killing vector is zero there. Physically, this region is very important because an 
outside observer can go through it and return back without any resistance. 
Penrose stated that the existence of an ergoregion around the rotating black hole 
provides a way to retract energy from the black hole \cite{Penrose:1971uk}. This 
process is generally called the Penrose process. We portrait different cases of 
the ergosphere by varying the values of the spin parameters $a$, $b$, and charge 
$q_e$ (cf. Fig.~\ref{ergo}). We find that there is an increase in area of the 
ergoregion due to the variation of charge $q_e$ as well as the spin parameters $a$ 
and $b$. The near extremal cases of the horizons can also be seen in Fig.~\ref{ergo}, 
where we precisely choose the values of the charge $q_e$ in such a way that the two 
horizons apparently nearly coinciding.
\begin{figure*}
\includegraphics[scale=0.415]{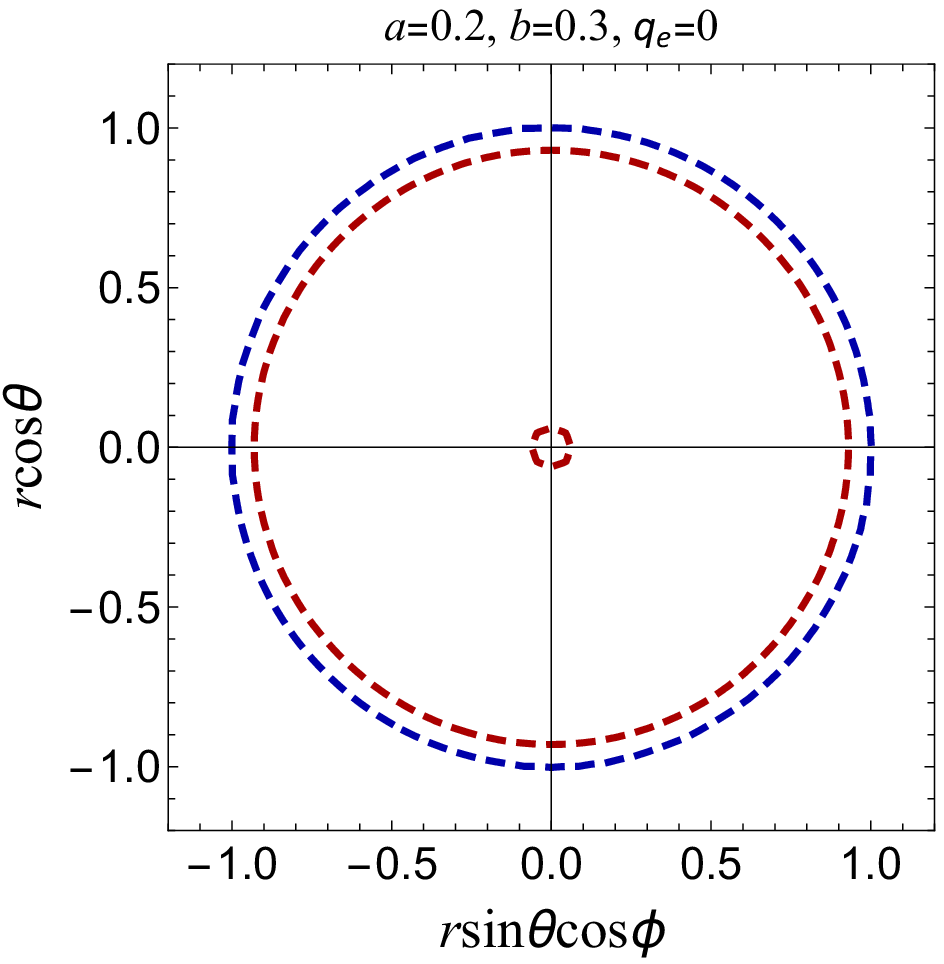}
\includegraphics[scale=0.415]{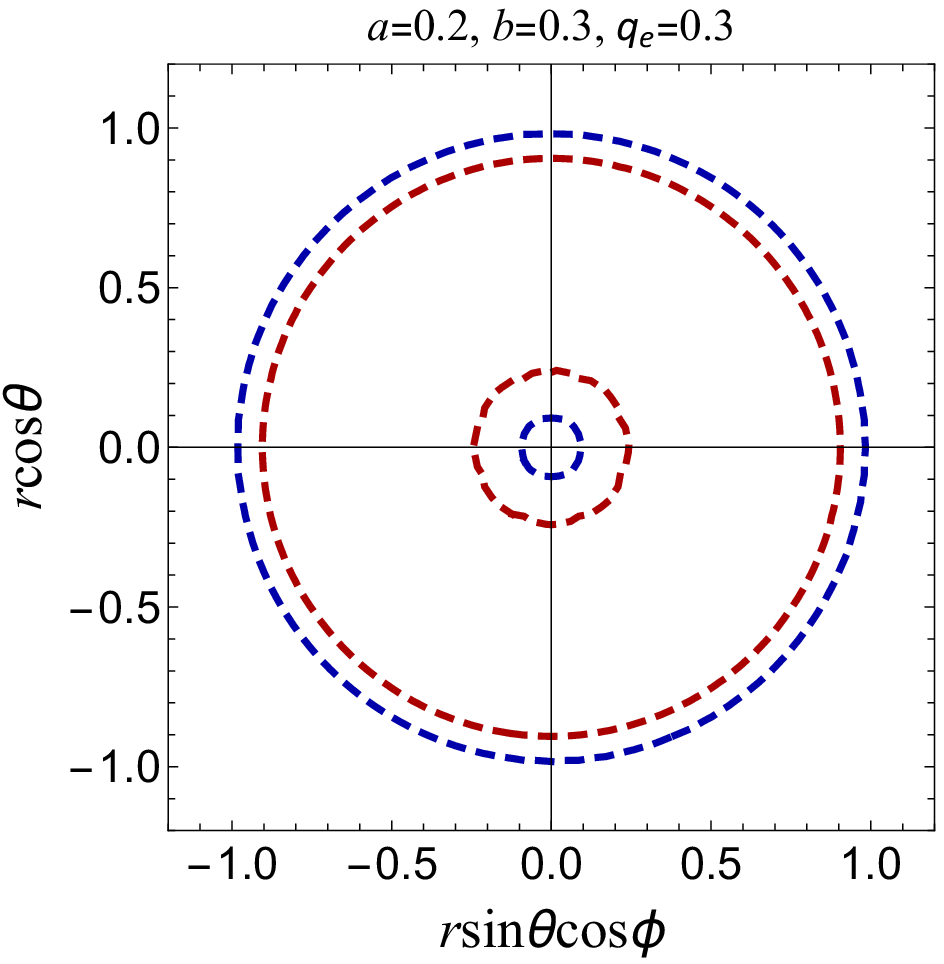}
\includegraphics[scale=0.415]{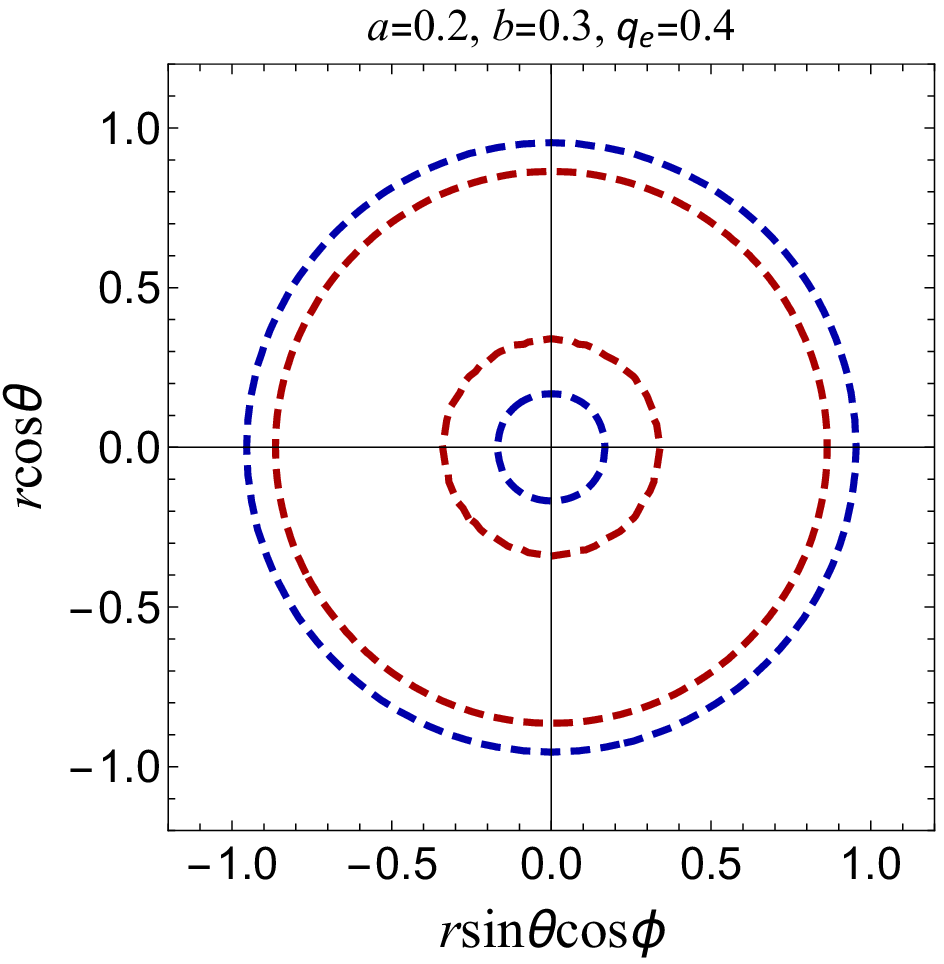}
\includegraphics[scale=0.415]{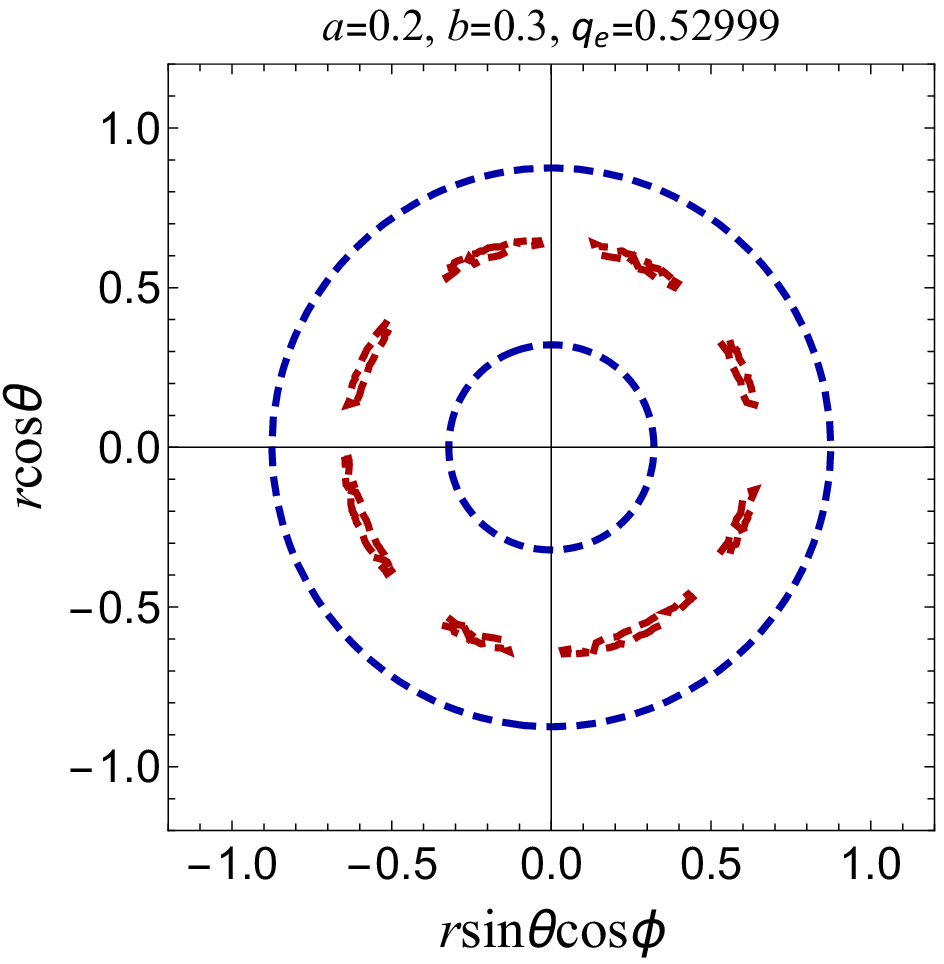}\\
\includegraphics[scale=0.415]{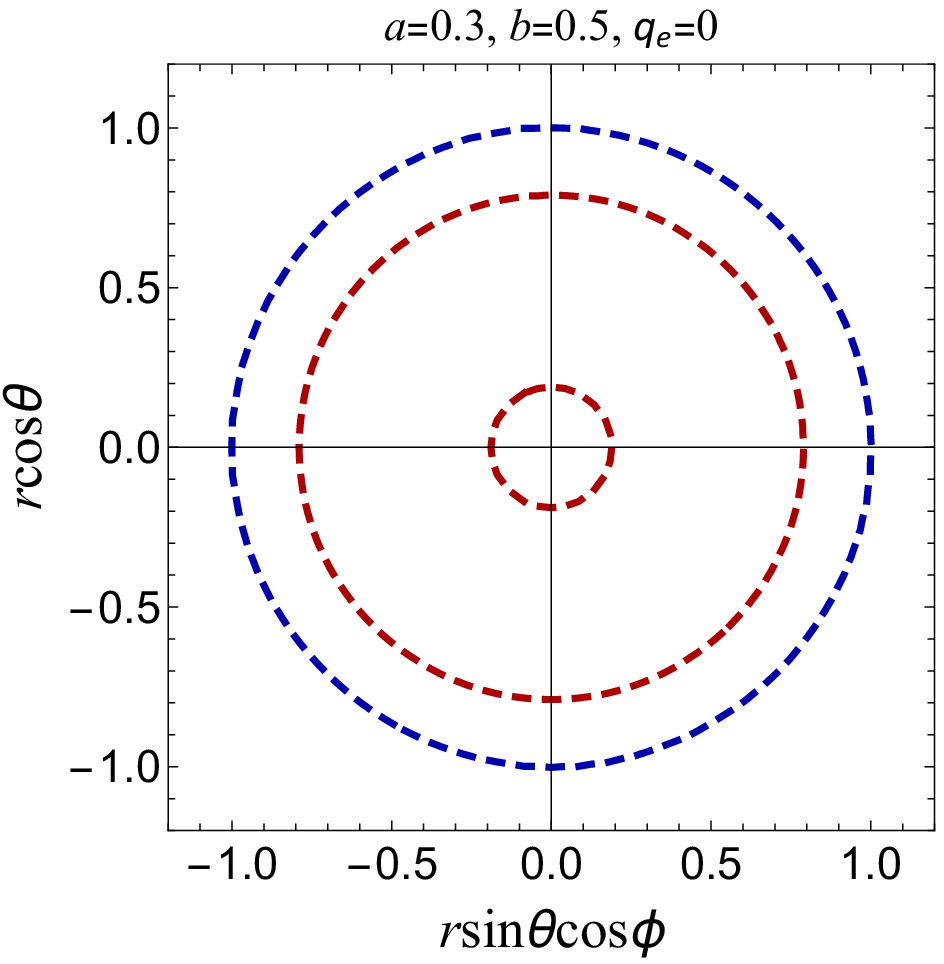}
\includegraphics[scale=0.415]{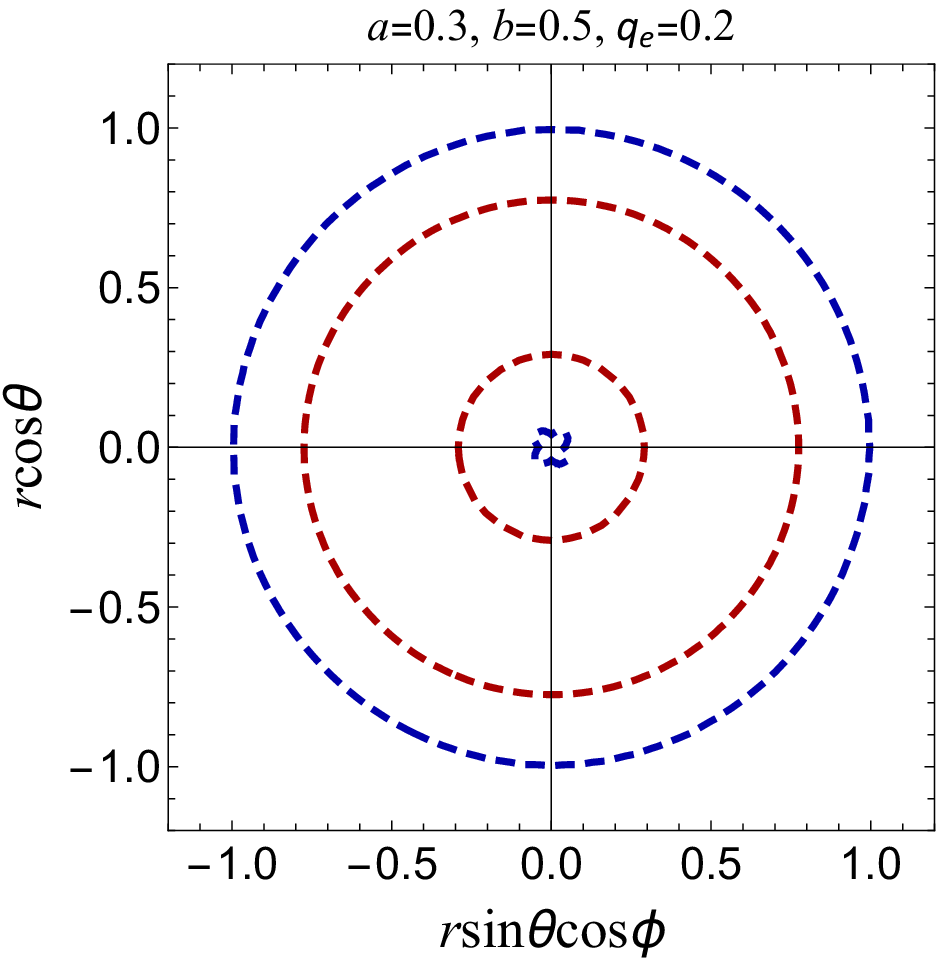}
\includegraphics[scale=0.415]{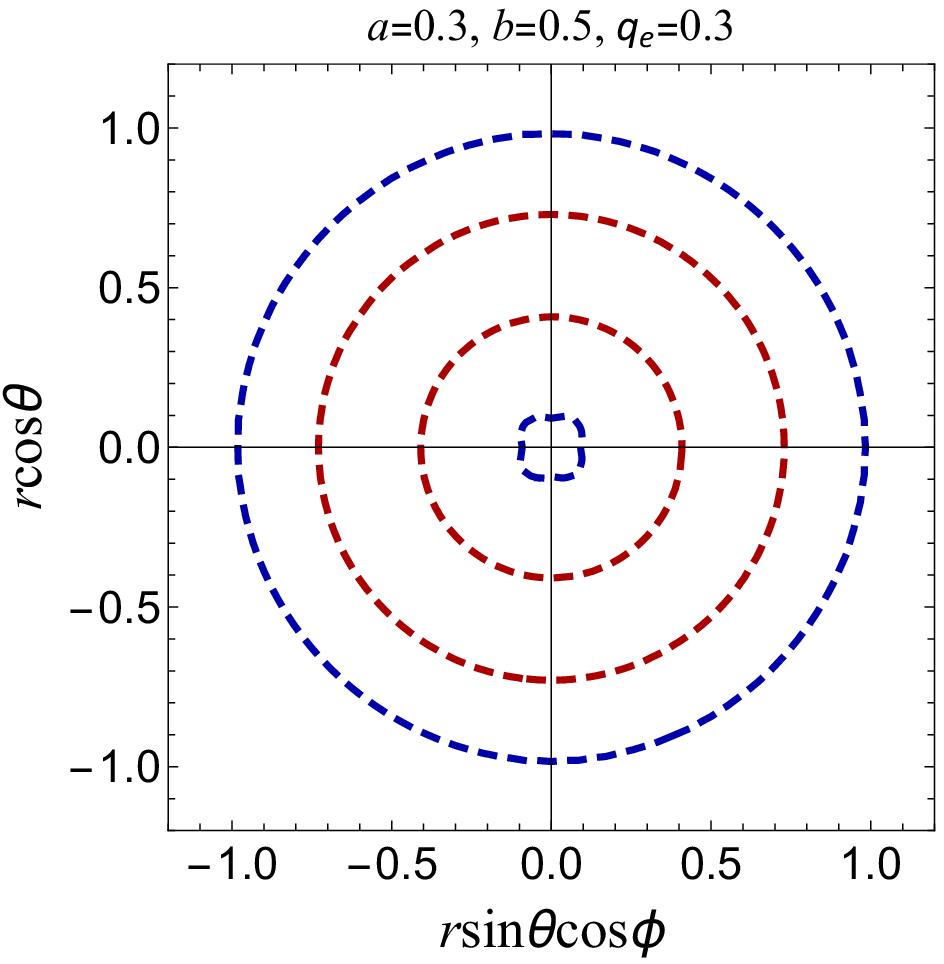}
\includegraphics[scale=0.415]{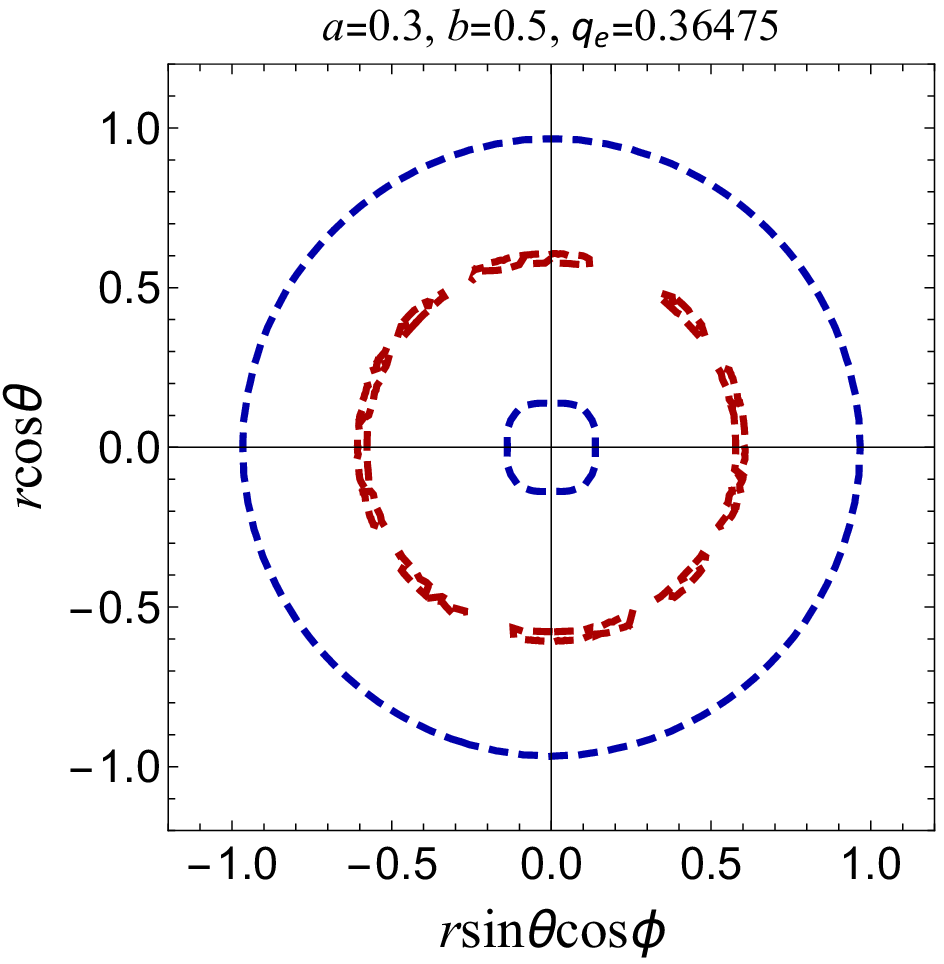}
\caption{\label{ergo} (color online). Plot showing the behavior of 
ergosphere of the rotating five-dimensional electrically charged Bardeen 
regular black hole.}
\end{figure*}
This particular illustration can be seen in right most column of Fig.~\ref{ergo}.

\section{Black hole thermodynamics}
\label{bhthd}
\begin{figure}
\includegraphics[scale=0.65]{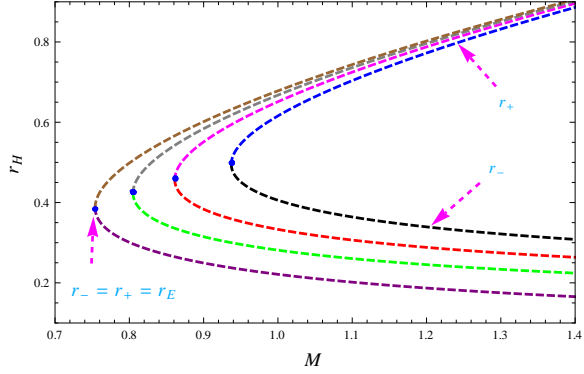}
\caption{\label{horizon} (color online). Plot showing the behavior of Cauchy 
horizon ($r_-$) and event horizon ($r_+$) of the rotating five-dimensional 
electrically charged Bardeen regular black hole.} 
\end{figure}

\begin{figure*}
\includegraphics[scale=0.65]{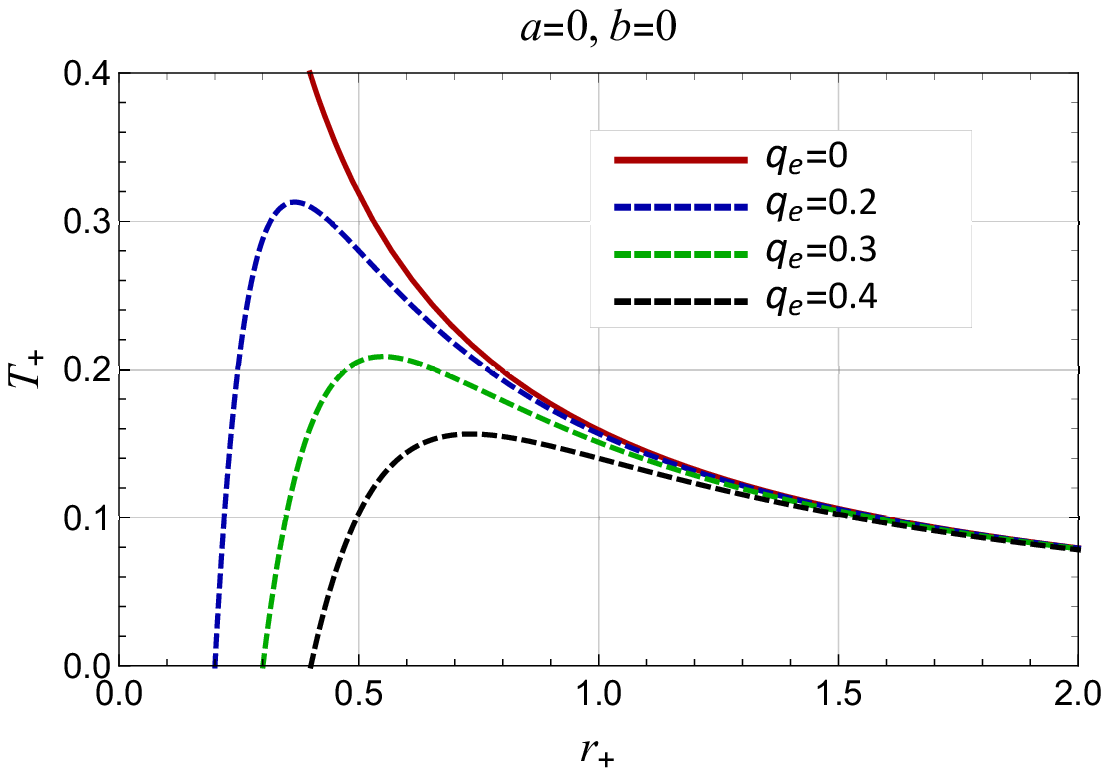}
\includegraphics[scale=0.65]{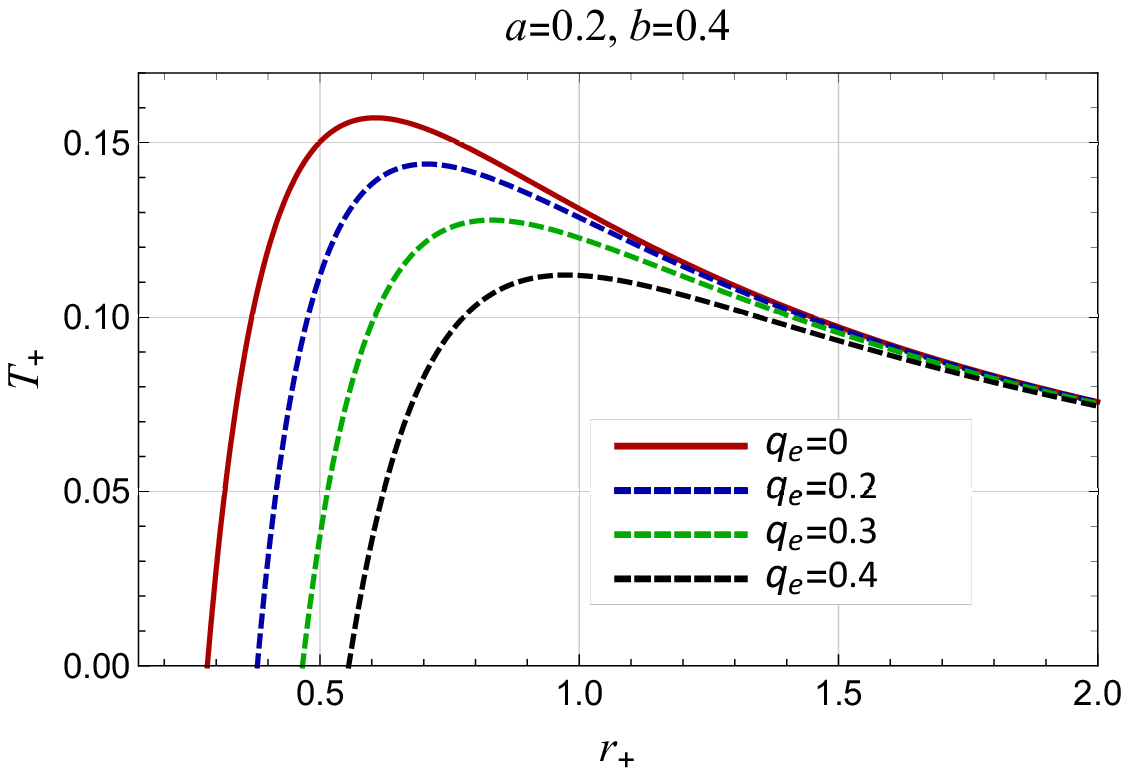}\\
\includegraphics[scale=0.65]{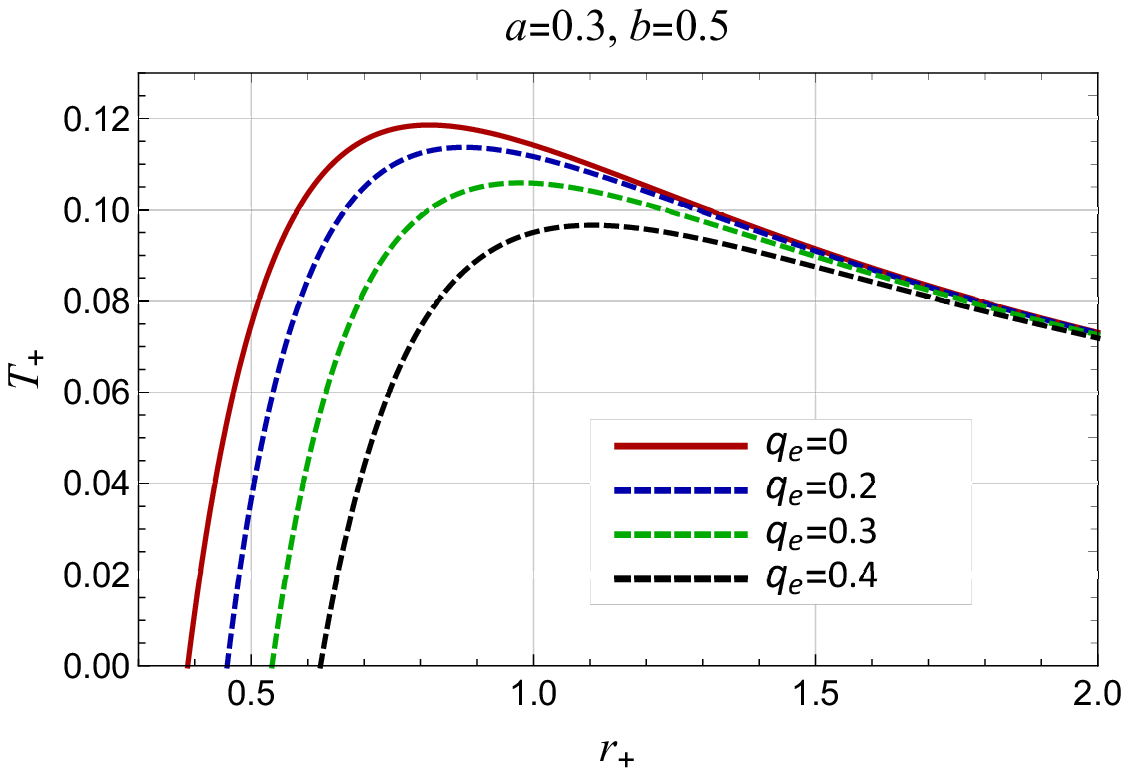}
\includegraphics[scale=0.65]{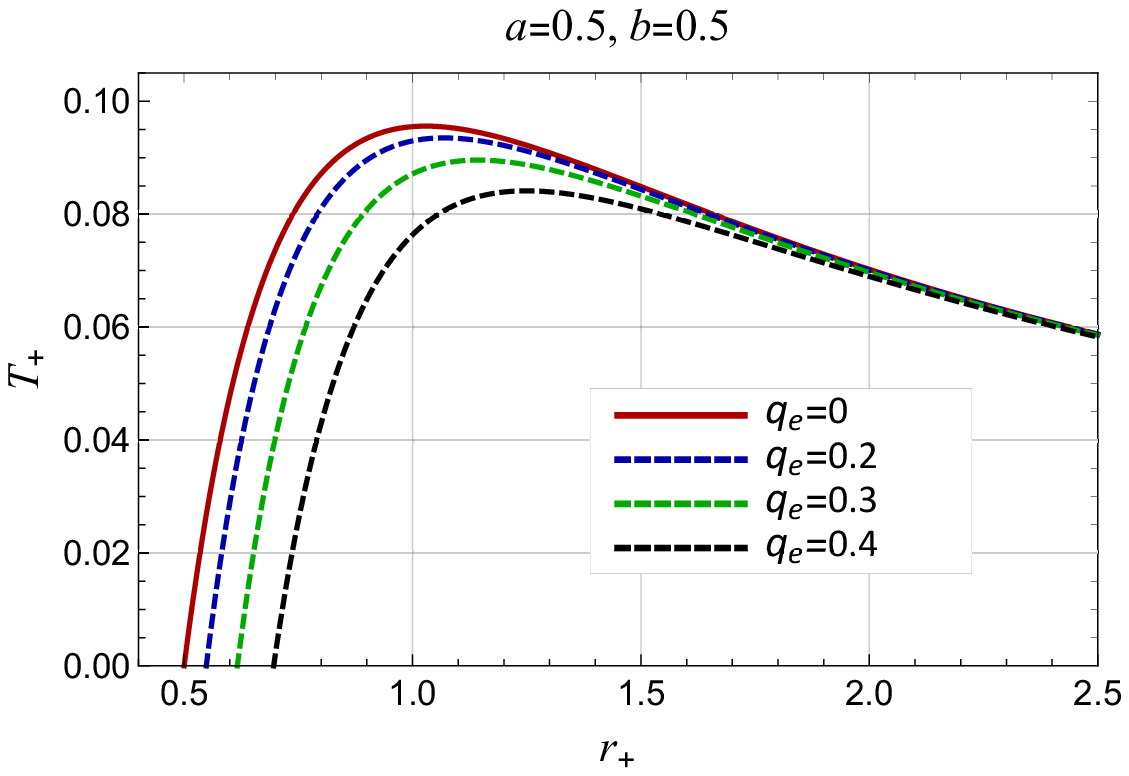}
\caption{\label{figT} (color online) Plot showing the behavior of Hawking 
temperature ($T_+$) vs event horizon radius ($r_+$) for rotating rotating 
five-dimensional electrically charged Bardeen regular black holes.}
\end{figure*}

In preceding section, we have addressed some important properties of the rotating 
five-dimensional electrically charged Bardeen regular black hole. Now we are going to 
discuss the black hole thermodynamics of this spacetime. The spacetime (\ref{metricf}) 
is equipped with one timelike Killing vector, $\partial/\partial{t}$ and two 
rotational Killing vectors, namely, $\partial/\partial{\phi}$ and 
$\partial/\partial{\psi}$ corresponding to the rotations about $\phi$-axis and 
$\psi$-axis, respectively. The null generators at the event horizon can be 
determined when three of them Killing vectors are combined together as follows
\begin{equation}\label{KF}
\ell = \frac{\partial}{\partial t} +\Omega_{\phi} \frac{\partial}{\partial \phi} 
+\Omega_{\psi} \frac{\partial}{\partial \psi},
\end{equation}
where $\Omega_\phi$ and $\Omega_{\psi}$ are the angular velocities of the spacetime 
corresponding to the $\phi$-axis and $\psi$-axis, respectively. Since $\ell$ becomes 
null on the event horizon at $r=r_+$ that leads to $\Delta(r_H)=0$, where 
$r_H=\lbrace r_-, r_+ \rbrace$ represents both the Cauchy horizon ($r_-$) and the event 
horizon ($r_+$) of the black holes. The angular velocities at the event horizon are 
given by
\begin{eqnarray}
\Omega_\phi = \frac{a (r_+^2 +b^2)}{(r_+^2 +a^2) (r_+^2 +b^2)},\quad 
\Omega_\psi = \frac{b (r_+^2 +a^2)}{(r_+^2 +a^2) (r_+^2 +b^2)}.
\end{eqnarray}
The mass of the rotating five-dimensional electrically charged Bardeen regular black 
holes can be computed as follows
\begin{eqnarray} \label{bhm}
M_+ = \frac{(r_+^2 +a^2) (r_+^2 +b^2) (r_+^3 +q_e^3)^{4/3}}{r_+^6}.
\end{eqnarray}
Figure~\ref{horizon} depicts the behavior of the horizon radius with the mass parameter 
for different values of the charge $q_e$, and the rotation parameters $a$ and $b$. 
It is clear that the two horizons coincides at $r_-=r_+=r_E$, corresponding to the 
extreme black hole configuration.    
\noindent
We further compute the Hawking temperature of the rotating five-dimensional 
electrically charged Bardeen regular black holes. The spacetime (\ref{metricf}) 
admits Killing field given in Eq.~(\ref{KF}) and correspondingly we have a conserved 
quantity associated with it. It is possible to construct a conserved quantity by 
using the Killing field such that
\begin{eqnarray}
\nabla^{\nu}\left(\ell^{\mu} \ell_{\mu}\right)=-2\kappa \ell^{\nu}, 
\end{eqnarray}
which in turn yields
\begin{eqnarray}
\kappa=\sqrt{-\frac{1}{2}\nabla_{\mu} \ell_{\nu} \nabla^\mu \ell^{\nu}}.
\end{eqnarray}
Here $\nabla^\nu$ represent covariant derivative and $\kappa$ is constant along 
the $\ell^{\mu}$ orbits which leads to the vanishing of Lie derivative of 
$\kappa$ along  $\ell^{\mu}$, i.e., $ \mathcal{L_\ell} \kappa=0 $. 
$\kappa$ is also known as the surface gravity which is proven to be constant over the 
event horizon. We can easily compute the surface gravity 
\begin{equation}
\kappa = \frac{r_+^7-q_e^3 r_+^4 -a^2 b^2 r_+^3 -2(a^2 + b^2)q_e^3 r_{+}^2 
    -3 a^2 b^2 q_e^3}{r_+ (r_+^2 +a^2) (r_+^2 +b^2) (r_+^3 +q_e^3)},
\end{equation}
The surface gravity is related to the Hawking temperature of the 
black holes via $T_+=\kappa/2\pi$. Hence, the Hawking temperature of the 
rotating five-dimensional electrically charged Bardeen regular black holes 
is given by
\begin{eqnarray}\label{temp}
   T_+ =\frac{r_+^7-q_e^3 r_+^4 -a^2 b^2 r_+^3 -2(a^2 + b^2)q_e^3 r_{+}^2 
   -3 a^2 b^2 q_e^3}{2 \pi r_+ (r_+^2 +a^2) (r_+^2 +b^2) (r_+^3 +q_e^3)}.
\end{eqnarray}
The black holes emit radiation when heated to the Hawking temperature. It is 
noticeable that equation (\ref{temp}) contains the Hawking temperatures for 
Meyers-Perry spacetime when $q_e=0$, the five-dimensional electrically charged 
Bardeen regular spacetime when $a=b=0$, and the Schwarzschild-Tagherlini 
spacetime when $q_e=a=b=0$ as special cases. The Hawking temperature does not 
diverge except for the $q_e=a=b=0$ special case and shows a finite peak 
(cf. Fig.~\ref{figT}). The typical behavior of the Hawking temperature as a function 
of the event horizon radius is depicted in Fig.~\ref{figT} for different values of 
the charge $q_e$, and the rotation parameters $a$ and $b$. These figures suggest that 
the regular black holes ($q_e\neq 0$) are colder than the singular black holes 
($q_e=0$). The Hawking temperature of the Schwarzshild-Tangherlini spacetime 
($T_+=1/2\pi r_+ $) shows the divergent behavior as $r_+\rightarrow 0$. 

\begin{figure*}
\includegraphics[scale=0.65]{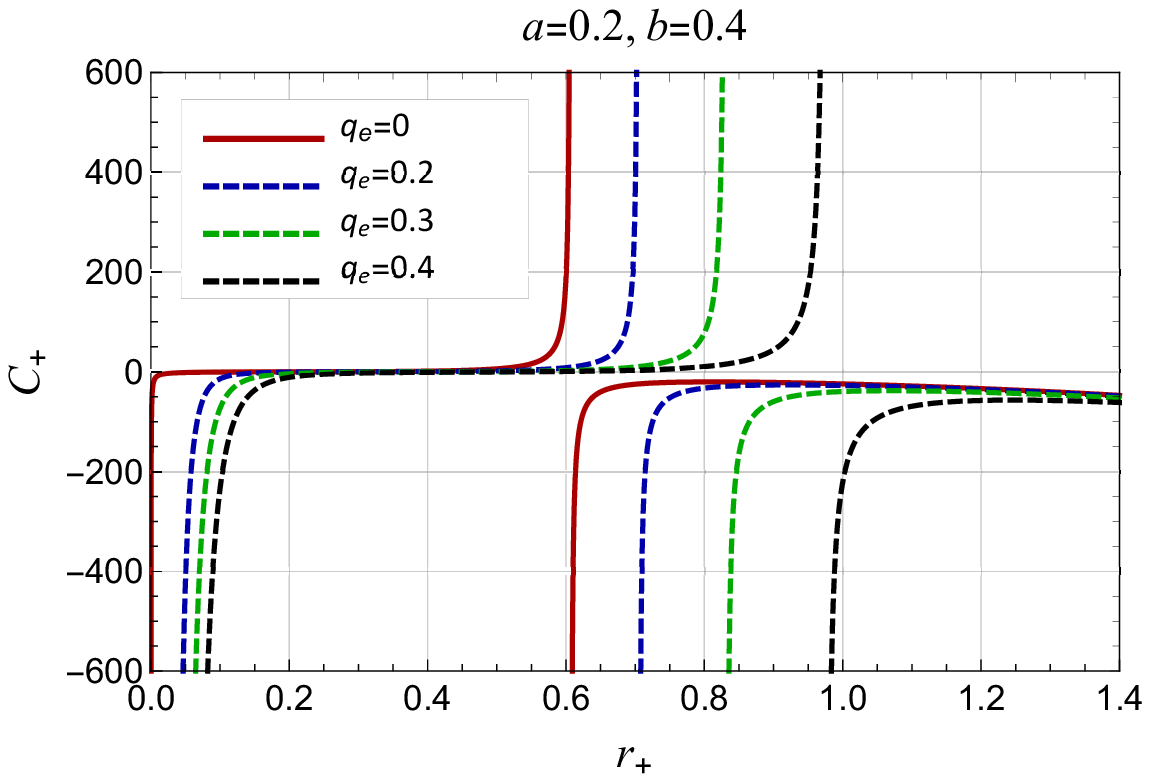}
\includegraphics[scale=0.65]{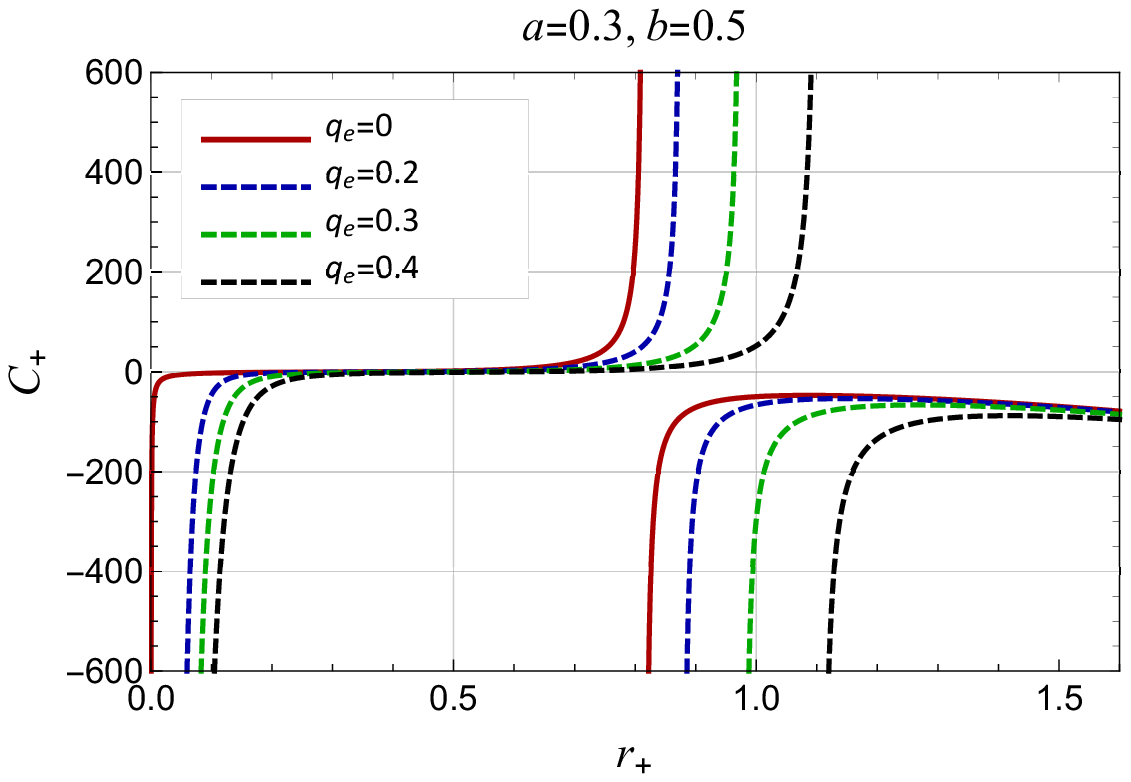}
\caption{\label{figC} (color online) Plot showing the behavior of heat 
capacity ($C_+$) vs event horizon radius ($r_+$) for rotating five-dimensional 
electrically charged Bardeen regular Bardeen black holes.}
\end{figure*}

Now the heat capacity of a black holes can be evaluated by using the following standard 
definition
\begin{eqnarray}\label{hcform}
C_+ = \left(\frac{\partial M_+}{\partial r_+}\right)
\left(\frac{\partial r_+}{\partial T_+}\right).
\end{eqnarray}
On using (\ref{bhm}) and (\ref{temp}) in (\ref{hcform}), the heat capacity for 
five-dimensional electrically charged Bardeen regular black holes turns out to be
\begin{eqnarray}\label{heatC}
C_+= \frac{4\pi (r_+^2+a^2)^2 (r_+^2+b^2)^2 (r_+^3 +q_e^3)^{7/3}
\lbrace r_+^7-q_e^3 r_+^4 -a^2 b^2 r_+^3 -2(a^2 + b^2)q_e^3 r_{+}^2 
-3 a^2 b^2 q_e^3 \rbrace}{r_+^5 \lbrace a^4 b^4 W +2q_e^3r_+^4(a^4+b^4)
(4r_+^3 +q_e^3) +8a^2b^2r_+^4 X +r_+^2 (r_+^2+a^2)Y +r_+^8 Z \rbrace}, 
\nonumber\\
\end{eqnarray}
where $W$, $X$, $Y$, and $Z$ are defined as follows
\begin{eqnarray}
W &=& r_+^6 +10q_e^3 r_+^3 +3q_e^6, \quad X = r_+^6 +6q_e^3 r_+^3 
+2q_e^6,\nonumber\\
Y &=& r_+^{10} +18q_e^3 r_+^7 +3a^2 b^2 r_+^6 +5q_e^6 r_+^4 +22a^2 b^2 q_e^3 
r_+^3 +7 a^2 b^2 q_e^6,
\nonumber\\
Z &=& -r_+^6 +6q_e^3 r_+^3 +q_e^6.
\end{eqnarray}
The expression has very complicated form which cannot be understand analytically. 
In order to see the nature of heat capacity, it will be convenient to plot the heat 
capacity versus event horizon radius by varying the charge $q_e$ as well as the 
rotation parameters (cf. Fig.~\ref{figC}).  
This is clear from Fig.~\ref{figC} that it is always possible for a suitable choice of 
the parameters to obtain the thermodynamically stable black holes. Figure~\ref{figC} 
reveals that there exists phase transitions in heat capacity. The heat capacity has 
two branches: i) the negative branch corresponds to the unstable phase, and ii) the 
positive branch corresponds to the stable phase for a suitable choice of the physical 
parameters.

\section{Conclusion}
\label{conclusion}
In this paper, we have discussed five-dimensional electrically charged Bardeen regular 
black holes spacetime by considering the coupling of nonlinear electrodynamics to 
general relativity. We have shown that the curvature invariants of the spacetime are 
not diverging at the origin ($r=0$) which indicates the regular behaviour of spacetime. 
Moreover, we have derived the rotating five-dimensional electrically charged Bardeen 
regular black hole spacetime by applying the Giampieri algorithm on the static 
five-dimensional electrically charged Bardeen regular spacetime. The Giampieri approach 
is much simpler than the NJA, since there is no need to do such tedious calculations. 
The rotating spacetime contains two distinct spin parameters $a$ and $b$ as well as a 
charge $q_e$ that provides a deviation, and differs it from the standard Myers-Perry 
spacetime. When we switch off the charge $q_e$, it reduces into the Myers-Perry black 
hole. In order to check the validity of the rotating spacetime, we have determined 
the nonlinear electrodynamics source. We have also computed the orthonormal basis by 
considering the locally nonrotating frames. By using these orthonormal basis, we have 
derived the components of the energy-momentum tensor, eventually discussed the energy 
conditions. We have discovered that there is violation of the weak energy condition 
in the rotating spacetime, but by a small amount.

A study of the different curvature invariants for the rotating spacetime reveals that 
it does not contain any spacetime singularity within it. One should note that the 
spacetime singularity has been removed from the spacetime through the nonlinear 
electrodynamics. This confirmation encourages us to study other properties of the 
spacetime. Therefore, the structures of the horizons, the static limit surfaces, and 
the ergosphere have been analysed in detail. While the horizons of the rotating 
spacetime are concerned, we have shown that there exists three different cases of 
particular interest, viz., two distinct roots corresponding to inner (Cauchy) horizon 
and outer (event) horizon, extremal configuration with degenerate horizons, and no root 
corresponding to no black hole solution. On a discussion regarding the ergosphere of 
the rotating five-dimensional electrically charged Bardeen regular spacetime, it has 
been found that the ergoregion increases with an increase in the magnitude of charge 
$q_e$ as well as spin parameters $a$ and $b$. We have further explored the 
thermodynamical behavior of the rotating five-dimensional electrically charged 
Bardeen regular spacetime by computing the Hawking temperature and the heat capacity. 
It has been found that there occurs phase transitions in heat capacity. In future work, 
we shall explore the phase space thermodynamics and critical behaviors, and critical 
exponents of the electrically charged Bardeen regular black holes. 

\acknowledgements
M.A. would like to thank University of KwaZulu-Natal and the National Research 
Foundation for financial support. MSA's research is supported by the ISIRD grant 9-252/2016/IITRPR/708. SDM acknowledges that this work is based on 
research supported by the South African Research Chair Initiative of the 
Department of Science and Technology and the National Research Foundation. 
We would like to thank the referees for useful comments and suggestions.

\end{document}